\documentclass[iop,revtex4,numberedappendix]{emulateapj}
\usepackage{graphicx}
\usepackage{natbib}
\usepackage{amssymb}
\usepackage{amsmath}
\usepackage{xcolor}
\usepackage[hyperindex,breaklinks,bookmarks=false]{hyperref}
\usepackage[figuresright]{rotating}

\newcommand{\oii}{[\ion{O}{2}] 3727\AA\;}
\newcommand{\oiii}{[\ion{O}{3}] 4959,5007\AA\;}
\newcommand{\hbeta}{H$\beta$}
\newcommand{\emni}{[\ion{N}{1}] 5199\AA\;}

\def\lesssim{\lower.5ex\hbox{$\; \buildrel < \over \sim \;$}}
\def\gtrsim{\lower.5ex\hbox{$\; \buildrel > \over \sim \;$}}

\shorttitle{Warm Ionized Gas in MASSIVE Early-type Galaxies}
\shortauthors{V. Pandya et al.}

\date{\today}

\begin{document}

\title{The MASSIVE Survey VI: The Spatial Distribution and Kinematics of Warm Ionized Gas in the Most Massive Local Early-type Galaxies}
  
\author {Viraj Pandya$^{1,2}$, Jenny E. Greene$^{1}$, Chung-Pei Ma$^3$, Melanie Veale$^3$, Irina Ene$^3$, Timothy A. Davis$^{4,5}$, John P. Blakeslee$^6$, Andy D. Goulding$^1$, Nicholas J. McConnell$^6$, Kristina Nyland$^7$, Jens Thomas$^8$}
\affil{$^{1}$Department of Astrophysical Sciences, Peyton Hall, Princeton University, Princeton, NJ 08540, USA}
\affil{$^{2}$UCO/Lick Observatory, Department of Astronomy and Astrophysics, University of California, Santa Cruz, CA 95064, USA}
\affil{$^{3}$Department of Astronomy, University of California, Berkeley, CA 94720, USA}
\affil{$^{4}$Centre for Astrophysics Research, University of Hertfordshire, Hatfield, Herts AL10 9AB, UK}
\affil{$^{5}$School of Physics \& Astronomy, Cardiff University, Queens Buildings, The Parade, Cardiff, CF24 3AA, UK}
\affil{$^6$Dominion Astrophysical Observatory, NRC Herzberg Institute of Astrophysics, Victoria, BC V9E 2E7, Canada}
\affil{$^7$National Radio Astronomy Observatory, Charlottesville, VA 22903, USA}
\affil{$^8$Max Planck-Institute for Extraterrestrial Physics, Giessenbachstr. 1, D-85741 Garching, Germany}
\email{viraj.pandya@ucsc.edu}  

\begin{abstract}
We present the first systematic investigation of the existence, spatial distribution, and kinematics of warm ionized gas as traced by the \oii emission line in 74 of the most massive galaxies in the local Universe. All of our galaxies have deep integral field spectroscopy from the volume- and magnitude-limited MASSIVE survey of early-type galaxies with stellar mass $\log(M_*/M_{\odot})>11.5$ ($M_K<-25.3$ mag) and distance $D<108$ Mpc. Of the 74 galaxies in our sample, we detect warm ionized gas in 28, which yields a global detection fraction of $38\pm6\%$ down to a typical [\ion{O}{2}] equivalent width limit of 2\AA. MASSIVE fast rotators are more likely to have gas than MASSIVE slow rotators with detection fractions of $80\pm10\%$ and $28\pm6\%$, respectively. The spatial extents span a wide range of radii ($0.6-18.2$ kpc; $0.1-4R_e$), and the gas morphologies are diverse, with $17/28\approx61\pm9\%$ being centrally concentrated, $8/28\approx29\pm9\%$ exhibiting clear rotation out to several kpc, and $3/28\approx11\pm6\%$ being extended but patchy. Three out of four fast rotators show kinematic alignment between the stars and gas, whereas the two slow rotators with robust kinematic measurements available exhibit kinematic misalignment. Our inferred warm ionized gas masses are roughly $\sim10^5M_{\odot}$. The emission line ratios and radial equivalent width profiles are generally consistent with excitation of the gas by the old underlying stellar population. We explore different gas origin scenarios for MASSIVE galaxies and find that a variety of physical processes are likely at play, including internal gas recycling, cooling out of the hot gaseous halo, and gas acquired via mergers. 
\end{abstract}


\maketitle

\section{Introduction}
Massive early-type galaxies typically have very old stellar populations. Given the lack of ongoing star formation, we expect a corresponding lack of an interstellar medium (ISM). However, a significant fraction of the massive early-type population has been observed to contain a complex and multi-phase ISM \citep[e.g.,][]{knapp85,phillips86,knapp96,goudfrooij99,sarzi06,sarzi10,davis11,davis16a}. Roughly $10^7-10^9M_{\odot}$ worth of molecular gas is detected in $\sim25\%$ of typical early-type galaxies \citep[][and references therein]{knapp96b,combes07,young11}. The warm ionized gas traces the dust in these galaxies quite well in terms of both mass fraction and morphology \citep[roughly $10^3-10^5M_{\odot}$; e.g.,][]{kim89,buson93,goudfrooij94}, and is often accompanied by the detection of molecular gas \citep[e.g.,][]{young11,davis11,alatalo13}. The relationship between the warm ionized and hot X-ray-emitting gas is far less clear \citep[e.g.,][]{sparks89,macchetto96,goudfrooij99}, and may well depend on the temperature profile of the X-ray-emitting gas \citep[e.g.,][]{mcdonald10,werner14}.

Despite more than thirty years of studying warm ionized gas in early-type galaxies \citep[e.g.,][]{caldwell84,demoulinulrich84,phillips86,buson93,goudfrooij94,zeilinger96,macchetto96,caon00,martel04,sarzi06,sarzi10}, we still do not have a clear understanding of the different origin and excitation mechanisms of the ionized gas. We do not know if the main sources are internal (e.g., stellar mass loss) or external (e.g., mergers). The total ISM masses that we infer are in general lower than predictions from stellar mass loss models, suggesting that the recycled gas is heated or expelled in a timely manner \citep[although our understanding of stellar mass loss is not complete; e.g., see][]{athey02,temi07,martini13,simonian16}. Given the typically low star formation rates of early-type galaxies, it is interesting to ask where the gas comes from, whether it is long-lived or recently accreted/cooled, and what excites the gas. 

One way to proceed is to ask whether the presence of ionized gas correlates with other properties of the galaxies. In particular, early-type galaxies are typically divided into two classes that are thought to be related to their formation history \citep[e.g.,][]{kormendy09,cappellari16}. Less massive early-type galaxies tend to be flattened, have younger stellar populations, and show regular rotation around the major axis \citep[fast rotators; e.g.,][]{emsellem07}. In contrast, the most massive early-type galaxies tend to be round and old and have very little net rotation (slow rotators). Slow rotators show a number of indirect signs of past merging activity, including kinematically decoupled components \citep[e.g.,][]{krajnovic11,naab14}. 

An excellent data set to address these questions is provided by the integral-field ATLAS$^{\rm 3D}$ survey \citep{cappellari11}, a volume-limited sample of early-type galaxies with $\log(M_*/M_{\odot})\gtrsim9.8$ ($M_K<-21.5$ mag) out to a distance of 42 Mpc. The ATLAS$^{\rm 3D}$ survey \citep[and its predecessor SAURON;][]{dezeeuw02} find a wide range of kinematic misalignment angles between the stars and gas in slow rotators, suggesting that in general the gas in these systems has been accreted \citep{sarzi06,davis11,sarzi13}. On the other hand, more than half of fast rotators are observed to host warm ionized gas kinematically aligned relative to the stars, which at least suggests that in a significant fraction of fast rotators, the gas may arise from internal processes such as stellar mass loss \citep[][]{sarzi06,davis11}, although kinematic alignment can occur in merger remnants as well \citep{vandevoort15}. Galaxies in richer environments are more likely to show kinematic alignment between the stars and gas compared to more isolated galaxies \citep{davis11}. In addition, the most massive fast rotators in ATLAS$^{\rm 3D}$, with $\log(M_*/M_{\odot})\gtrsim10.9$, uniformly show kinematic alignment between the gas and stars. However, because of the small volume limit of the ATLAS$^{\rm 3D}$ survey, they have poor coverage of the most massive early-type galaxies \citep[only 14\% of their galaxies are slow rotators;][]{emsellem11}.

The MASSIVE survey \citep{ma14} of early-type galaxies within 108 Mpc (including the Coma cluster) provides an ideal sample to study the warm ISM in galaxies that are a factor of ten more massive than the typical galaxies included in previous systematic surveys such as ATLAS$^{\rm 3D}$, and to probe a wider range of slow rotator properties. Since the galaxies we will consider in this paper all have $\log(M_*/M_{\odot})>11.5$ ($M_K<-25.3$ mag), our sample is dominated by slow rotators \citep{veale16}. Using integral field unit (IFU) spectroscopy obtained as part of the MASSIVE survey \citep{ma14}, we will present the first systematic investigation of the existence, spatial distribution, and kinematics of warm ionized gas in the most massive galaxies in the local Universe.

The structure of this paper is as follows. In \autoref{sec:sample}, we present our sample of MASSIVE galaxies. In \autoref{sec:methods}, we describe our methods for detecting warm ionized gas with our IFU observations. In \autoref{sec:results}, we present the detection fractions, spatial extent, line ratios and kinematics of the gas. In \autoref{sec:discussion}, we consider the implications of our results for the origins of excitation mechanisms of the gas. We summarize our results and discuss future work in \autoref{sec:summary}. 

\section{Sample and Data Description}\label{sec:sample}
Since the MASSIVE sample is thoroughly described in \citet{ma14}, here we will only briefly review the selection criteria. The MASSIVE survey targets the most massive local early-type galaxies by requiring $M_K<-25.3$ mag \citep[based on the 2MASS Redshift Survey;][]{huchra12} and applying a morphological cut to remove large spiral and interacting galaxies \citep[based on the Hyperleda database;][]{paturel03}. The survey is volume-limited out to a distance of 108 Mpc, which results in a factor of ten greater volume compared to previous similar IFU surveys \citep[e.g., ATLAS$^{\rm 3D}$;][]{cappellari11} and allows for the inclusion of many more massive early-type galaxies such as those in the Coma cluster.\footnote{We have removed NGC 7681 from our sample even though it hosts warm ionized gas near its nucleus because it is actually a close pair and each individual galaxy is unlikely to make it into the MASSIVE sample \citep[see also][]{veale16}.} The full MASSIVE sample consists of 116 galaxies, but the sample size for this paper is 74 galaxies (these constitute, with a few exceptions, the brighter two-thirds of the survey with $M_K<-25.5$ mag). 

This paper is based on integral field spectra obtained with the Mitchell Spectrograph (formerly VIRUS-P) on the 2.7m Harlan J. Smith Telescope at McDonald Observatory. The Mitchell IFU spectrograph has a $107^{\prime \prime}\times107^{\prime \prime}$ field of view with 246 $4^{\prime \prime}$-diameter fibers per pointing. The spectra cover the wavelength range from $\sim3500$\AA\ to $\sim5800$\AA. Data reduction details are given in \citet{greene12,greene15}, but see also \citet{veale16} for additional details. We generally have three dithered pointings for a total of 738 spectra per galaxy. We work with the single-fiber IFU spectra as well as a radially and azimuthally binned IFU dataset wherein fibers are grouped together to ensure continuum $S/N\geq20$ in each spatial bin. Fibers with spectra that have abnormally high median fluxes, exceptionally low continuum S/N, or are near foreground stars are masked. The spatial bins are constructed from the individual fibers (remaining after masking) from the center outward, leaving central fibers that already meet the S/N threshold unbinned. Beginning at the first fiber below the S/N threshold, we divide fibers into circular radial bins and choose azimuthal bins (of equal size within each annulus but not across annuli) that minimize the aspect ratio to avoid thin rings or wedges. The radius of each annulus is then increased, from center outward (reducing the number of azimuthal bins when needed to maintain the proper aspect ratio), until the S/N threshold is reached \citep[for more details, see][]{veale16}.

This paper represents part of our comprehensive effort to study gas of all phases in the MASSIVE sample, including using CO to trace molecular gas \citep[][]{davis16a} as well as hot X-ray gas \citep[][]{goulding16}.

\section{Methods}\label{sec:methods}
Here we describe our methods for detecting warm ionized gas emission in MASSIVE galaxies.

\subsection{Stellar Continuum Subtraction}
For robust detection of faint lines, it is crucial that we model and subtract the stellar continuum. We must also ensure that each spectrum is at rest with respect to the stars. For each IFU spectrum, we therefore fit the stellar continuum using the penalized pixel-fitting method \citep[pPXF;][]{cappellari04}. We only solve for the first two moments ($V_{\rm stars},\sigma_{\rm stars}$) of the full Gauss-Hermite decomposition \citep[see][]{vdmarel93,gerhard93}. We adopt the \citet{bc03} library of stellar population templates that span a grid of ages (5 Myr to 12 Gyr) and metallicities ($Z=0.004$, 0.02, and 0.05). To help account for flux calibration and sky subtraction errors, we fit with sixth order additive and sixth order multiplicative polynomials. We found that our \hbeta, [\ion{O}{2}] and [\ion{O}{3}] equivalent widths did not change significantly if we instead used first-order additive and twelfth-order multiplicative polynomials. This is an important check because additive polynomials, unlike multiplicative polynomials, can change the equivalent widths of stellar absorption lines \citep[e.g., see][]{koleva09}.

Each best-fit composite template spectrum is subtracted from its corresponding observed spectrum. The residual spectrum is used to measure the emission lines. Note that \citet{sarzi06} advocate an iterative approach; we do not implement such an algorithm here. We verified that the RMS of the residual spectrum is roughly comparable to the pixel-by-pixel uncertainties of the original spectrum, which means that we can use the RMS of the local continuum around an emission line to define the significance of our fits (see next subsection). All gas and stellar velocities presented in this paper are relative to the galaxy systemic velocity \citep[see also][]{veale16}.

\subsection{Emission Line Detection Algorithm}\label{sec:emlinedet}
Since our spectra span the wavelength range from $\sim3500$\AA\ to $\sim5800$\AA, we constrain the presence of the \oii, \oiii, \hbeta\;and \emni emission lines as follows. We begin by fitting [\ion{O}{2}] with a single Gaussian plus zero-slope additive continuum offset. The Gaussian is constrained to have positive amplitude, $1.0\rm\AA<\sigma_{\rm gas}<4.0\rm\AA$ (80 to 320 km s$^{-1}$ velocity dispersion), and a maximum central wavelength of $\mu=3727\rm\AA\pm10\rm\AA$. The constraint on $\sigma_{\rm gas}$ prevents very narrow Gaussians from being fit to noise spikes and very wide Gaussians from being fit to broad continuum residuals. The constraint on $\mu$ ensures that we allow only a reasonable range of gas velocities ($\pm800$ km s$^{-1}$). Although it is of course possible that emission lines can exist with velocities beyond that which our range of $\mu$ allows, we have confirmed through visual inspection of the spectra that no such cases exist in our sample.

We define the significance of a fit as the ratio of the line amplitude to the RMS of the local continuum residuals (i.e., after best-fit template subtraction), as in \citet{sarzi06}: $\mathcal{S}\equiv A/N$. [\ion{O}{2}] fits with $\mathcal{S}\geq3$ are considered to be significant. For fits with $\mathcal{S}<3$, we derive the $3\sigma$ upper limit on the integrated flux of [\ion{O}{2}] using the known relation between Gaussian amplitude, width, and integrated area while assuming $A=3N$ and $\sigma_{\rm gas}$=200 km s$^{-1}$.

The fit to [\ion{O}{3}] is similar to that for [\ion{O}{2}], but we fix $V_{\rm gas}$ and $\sigma_{\rm gas}$ to the values derived for [\ion{O}{2}]. H$\beta$ is simultaneously fit with the [\ion{O}{3}] doublet by fixing its kinematics to those of [\ion{O}{2}]. In cases where [\ion{O}{2}] is not detected, we allow $V_{\rm gas}$ and $\sigma_{\rm gas}$ to be free parameters of the shared line profile for [\ion{O}{3}] and \hbeta\;(with similar constraints as for [\ion{O}{2}]). We require $\mathcal{S}\geq3$ for significant [\ion{O}{3}] detections, but $\mathcal{S}\geq2.5$ for H$\beta$ fits. It is more difficult to constrain the presence of H$\beta$ because of the strong stellar absorption, but after we subtract the expected absorption contribution from stars based on our best-fit pPXF composite templates, excess H$\beta$ emission can be revealed. When necessary, upper limits on [\ion{O}{3}] and H$\beta$ are derived similarly to what would be done for [\ion{O}{2}].

The [\ion{N}{1}] emission line falls in the Mg$b$ wavelength region where the underlying stellar continuum is only poorly modeled by current stellar population templates, as described by \citet{sarzi06}. Therefore, we require more stringent significance criteria for it, following \citet{sarzi06,sarzi10}. Both [\ion{O}{3}] and H$\beta$ must be significantly detected in order for the [\ion{N}{1}] fit to be trusted. If both [\ion{O}{3}] and H$\beta$ are indeed significantly detected, and [\ion{N}{1}] itself has $\mathcal{S}\geq4$, then [\ion{N}{1}] is considered to be significantly detected. Otherwise, an upper limit is derived for [\ion{N}{1}] as described above. We detect [\ion{N}{1}] only in three spectra, and these are the fibers closest to the photometric centers of NGC 0708, NGC 1167, and NGC 1453. 

The resolving power of our spectra is $R\approx830$ (roughly 150 km s$^{-1}$) near [\ion{O}{2}], and $R\approx1060$ (roughly 120 km s$^{-1}$) near \hbeta, [\ion{O}{3}] and [\ion{N}{1}]. We subtract these instrumental broadening values in quadrature from our measured line widths. However, the procedure for [\ion{O}{2}] is a bit more involved because it is actually an unresolved doublet with a line separation of only 2.8\AA.\footnote{[\ion{N}{1}] is also an unresolved doublet with a line separation of only 2.4\AA, and in principle we should also correct its measured line width for blending. However, [\ion{N}{1}] is in a poorly modeled part of the stellar continuum and we rarely detect it, as described above. Therefore, we only correct the measured line width of [\ion{N}{1}] for instrumental broadening.} Before correcting for instrumental broadening of [\ion{O}{2}], we apply an empirical deblending calibration to its measured line width, which is derived as follows \citep[see also][]{salviander07}. For a grid of input velocity dispersions (50 to 300 km s$^{-1}$), we create and sum two Gaussians with the same amplitude, fixed line separation, and input line width. We fix the amplitude ratio [\ion{O}{2}]$\lambda3728.8$/[\ion{O}{2}]$\lambda3726.0$ to one because this corresponds to electron number density $n_e=100$ cm$^{-3}$, which is sensible for the diffuse ISM of early-type galaxies where the actual densities are much lower than the critical density for both doublet lines, $n_e\sim10^3$ cm$^{-3}$ \citep[see, e.g.,][]{osterbrock89}. We then fit a single Gaussian to each of the summed Gaussians and obtain a one-to-one mapping between intrinsic line width and blended line width. We use a quadratic polynomial fit to the empirical mapping to convert measured line widths to intrinsic line widths. If the deblending calibration results in an intrinsic line width that is below our resolution limit, we simply set the line width to the resolution limit; otherwise, we further subtract the instrumental broadening in quadrature.

\begin{figure*} 
\begin{center}
\includegraphics[width=0.9\hsize]{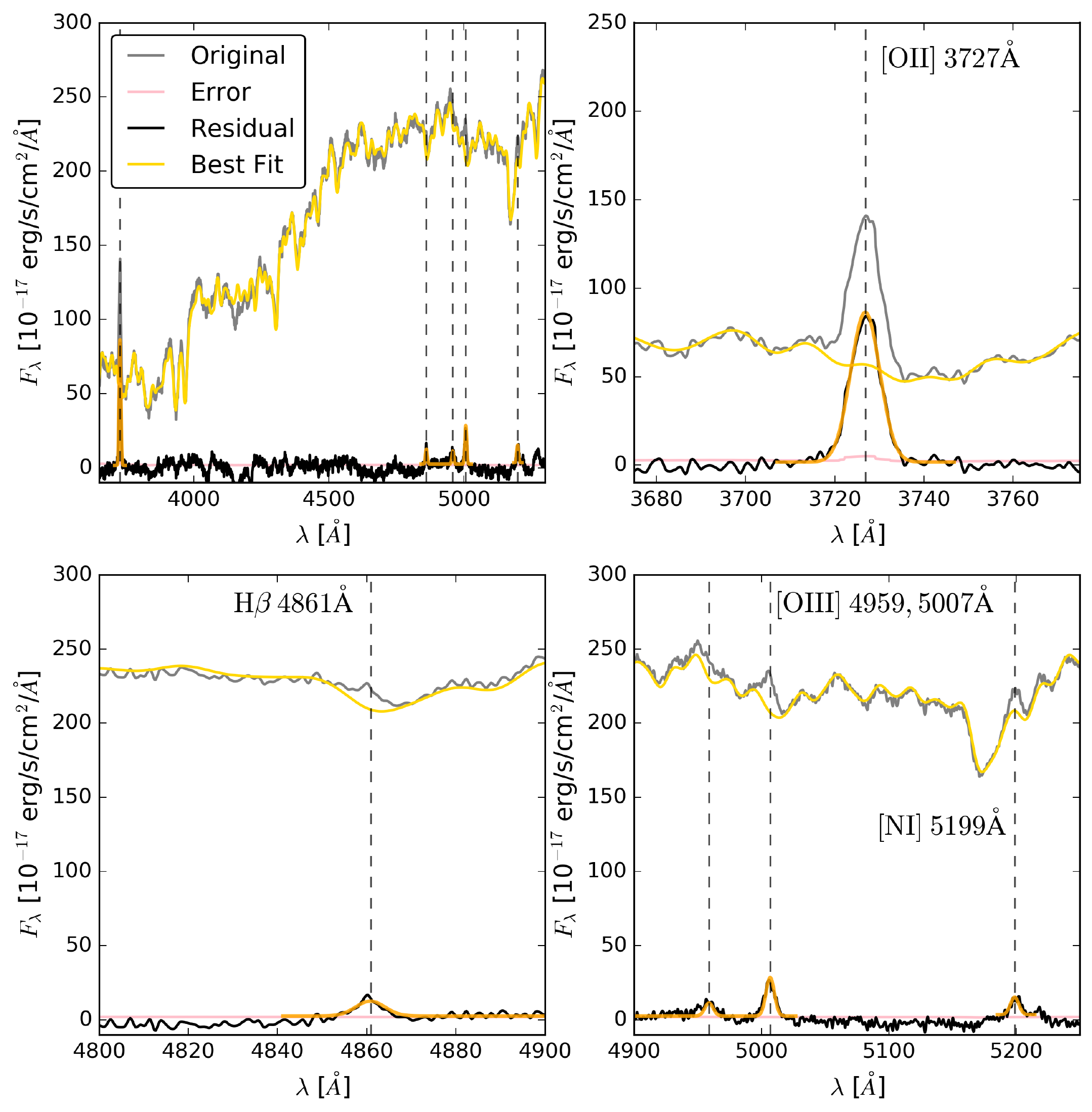}
\end{center}
\caption{Example of detecting [\ion{O}{2}], [\ion{O}{3}], H$\beta$ and [\ion{N}{1}] emission lines in the central fiber of NGC 1453. The original spectrum is shown in gray, best-fit pPXF composite template in gold, residual spectrum in black, error spectrum in pink, and best-fit Gaussians to emission lines in orange. Vertical dashed lines mark the rest-frame central wavelengths of the various emission lines.}
\label{fig:lines}
\end{figure*}

\begin{figure*} 
\begin{center}
\includegraphics[width=\hsize]{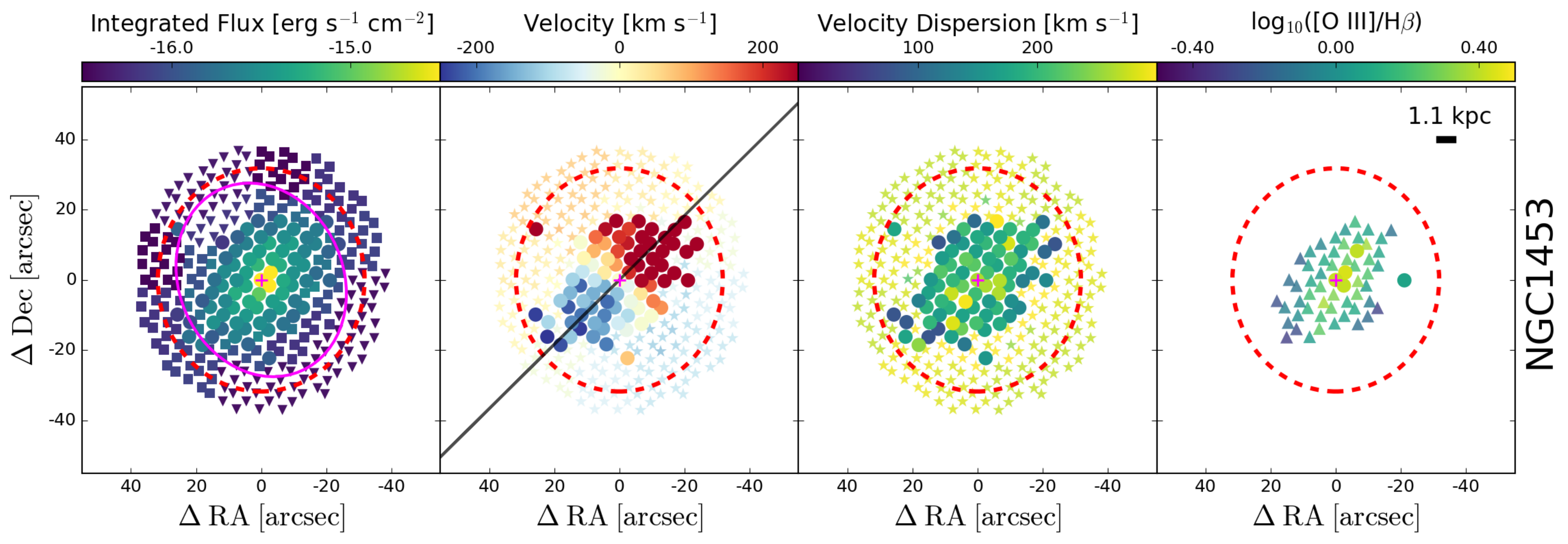}
\end{center}
\caption{Two-dimensional maps of various quantities for NGC 1453. From left to right: [\ion{O}{2}] integrated flux, gas and stellar velocity, gas and stellar velocity dispersion, and [\ion{O}{3}]/H$\beta$ integrated flux ratio. The pink cross in each subpanel marks the photometric center of the galaxy. The pink ellipse in the integrated flux subplot shows the stellar effective radius, photometric position angle, and axis ratio, whereas the red dashed circle in each panel marks the maximum radial extent of the gas, $R_{\rm gas}$. The black line in the velocity subplot shows the kinematic position angle of the rotating gas. In the velocity and velocity dispersion subplots, the star marker symbols in the background represent the velocities and velocity dispersions of the stars in each fiber as derived with pPXF. In all panels, circle marker symbols represent measurements from the individual fibers and squares represent measurements from the bins (shown where the single-fiber measurements are insignificant or have large uncertainties). In the integrated flux subplot, downward facing triangles represent upper limits, whereas in the excitation ratio subplot, upward facing triangles represent lower limits (shown where [\ion{O}{3}] is detected but H$\beta$ is not, leading to an H$\beta$ upper limit in the denominator). The number of kpc corresponding to four arcsec at the redshift of NGC 1453 is written in the rightmost subplot. Similar two-dimensional maps for the other 27 galaxies are shown in Appendix \ref{sec:maps2d}.}
\label{fig:portrait1453}
\end{figure*}

\autoref{fig:lines} shows the central fiber spectrum of NGC 1453 in which [\ion{O}{2}], [\ion{O}{3}], H$\beta$ and [\ion{N}{1}] are all significantly detected. \autoref{fig:portrait1453} shows example two-dimensional maps of various quantities for NGC 1453. 

\subsection{Sensitivity Limits}\label{sec:sens}
In Appendix \ref{sec:completeness}, we discuss simulations that assess our integrated flux completeness limits and uncertainties for [\ion{O}{2}], H$\beta$ and [\ion{O}{3}]. Here we will only briefly comment on the results of those simulations. Uncertainties on the measurements of detected emission lines (notably the integrated flux) were assumed to be $10\%$ in all cases based on the simulations. We can reliably detect the [\ion{O}{2}] emission line down to a minimum integrated flux of $\sim3\times10^{-16}$ erg s$^{-1}$ cm$^{-2}$.

We reach an EW limit of $\sim2$\AA\ and $\sim0.5$\AA\ for [\ion{O}{2}] and [\ion{O}{3}], respectively, in the central regions of galaxies where we work with individual fibers. In the binned data, we reach nearly an order of magnitude deeper in EW sensitivity, although stacking only works well when the gas fills much of a bin. Given that the stellar continuum level is three to four times fainter near [\ion{O}{2}] than near [\ion{O}{3}], we have comparable detection limits for the two lines in terms of integrated flux. The higher EW detection limit of [\ion{O}{2}] is due to the lower continuum S/N at blue wavelengths. While we use [\ion{O}{2}] as our primary detection line, none of our conclusions would change had we used [\ion{O}{3}] instead.

Since we will compare our results to those of the SAURON \citep{dezeeuw02} and ATLAS$^{\rm 3D}$ \citep{cappellari11} surveys, it is useful to mention their sensitivities. They both use the \texttt{SAURON} instrument \citep{bacon01} and utilize [\ion{O}{3}] for line detection ([\ion{O}{2}] is not in their observing band). The SAURON survey quotes limits in [\ion{O}{3}] and H$\beta$ of $\sim0.3$\AA\ and $\sim0.2$\AA, respectively \citep{sarzi06}. 

\section{Warm Ionized Gas Properties}\label{sec:results}
\subsection{Detection Fractions}\label{sec:detfrac} 
With the emission line measurements in hand, we now address whether a galaxy hosts warm ionized gas using [\ion{O}{2}] as our primary tracer. This is complicated by false positive [\ion{O}{2}] detections that can crop up in the outer parts of an IFU dataset (typically at the $\lesssim10\%$ level). We thus take two approaches to identify which galaxies host warm ionized gas. First, for the purpose of assigning a detection flag to each galaxy, we apply an additional [\ion{O}{2}]-local continuum $S/N\geq3$ cut. This additional $S/N$ cut removes most fibers within an IFU dataset except those within $\sim15$ arcsec of the photometric center, which generally have [\ion{O}{2}]-local continuum $S/N\gtrsim10$ and [\ion{O}{3}]-local continuum $S/N\gtrsim100$. The much higher local continuum $S/N$ ratios of these central fibers drastically reduces the chance of identifying noisy residual spikes as false positive [\ion{O}{2}] detections. 

If, after applying this continuum $S/N$ cut, there are $\geq3$ fibers with [\ion{O}{2}] detections remaining in the IFU data, then the galaxy is flagged as hosting warm ionized gas. If there are no fibers with [\ion{O}{2}] detections, then the galaxy is flagged as hosting no detectable warm ionized gas. In cases with only one or two fibers with detections, we find that only fibers close to the center and with $\mathcal{S}>3.5$ are reliable (in such cases we flag the galaxy as hosting a nuclear warm ionized gas reservoir). Although one may worry that these criteria discard galaxies that only have off-center [\ion{O}{2}] emission, we verified through visual inspection that there are no such cases in our sample.

The second complementary approach involves using the binned IFU spectra \citep[see \autoref{sec:sample} and][]{veale16}, which afford us significantly greater continuum $S/N$ at large galactocentric radii. In our outermost bins, we typically have [\ion{O}{2}]-local continuum $S/N>4$ and [\ion{O}{3}]-local continuum $S/N>30$.
 If the gas is uniformly distributed rather than patchy, binning can increase the significance of a detection. In every case where [\ion{O}{2}] is detected in our binned spectra with S/N $\sim20$ at large galactocentric radii, we also detect [\ion{O}{2}] in the central fibers. When individual fibers at large galactocentric radii do not reveal significant emission, sometimes the corresponding binned IFU spectra do reveal real emission.

We find that $28/74\approx38\pm6\%$ of the MASSIVE galaxies in our sample host a warm ionized medium.\footnote{Unless otherwise noted, the uncertainties that accompany all detection fractions are computed assuming a binomial distribution: $\delta f = \sqrt{f(1-f)/N}$, where $f$ and $N$ are the detection fraction and sample size, respectively.} The properties of these 28 galaxies are given in \autoref{tab:det}, ordered according to whether the gas is extended and rotating, extended and non-rotating, or centrally concentrated. Two-dimensional maps of various quantities for these galaxies are shown in \autoref{sec:maps2d} using the same order as in \autoref{tab:det} (NGC 1453 is shown separately in \autoref{fig:portrait1453}).  Most of these galaxies host warm ionized gas that is detected only in the central fibers: $17/28=61\pm9\%$. In $3/28\approx11\pm6\%$ galaxies, the warm ionized medium is spatially extended but patchy and non-rotating, and in $8/28\approx29\pm9\%$ of the MASSIVE galaxies, the warm ionized gas is spatially extended and shows regular rotation. If we split the full MASSIVE sample by stellar kinematic classification, then we find that $12/15=80\pm10\%$ of MASSIVE fast rotators host warm ionized gas compared to only $16/58=28\pm6\%$ MASSIVE slow rotators. A two-proportion z-test confirms that the detection fractions for MASSIVE slow and fast rotators are indeed statistically significantly different, with a z-score of 3.7 corresponding to a p-value $\ll0.01$. The distance and stellar mass distributions of MASSIVE fast and slow rotators are not significantly different from each other, as confirmed by non-parametric statistical tests. This means that the significantly different detection fractions for the two kinematic classes cannot be attributed to an underlying bias in their distance and stellar mass distributions.

\begin{figure*} 
\begin{center}
\includegraphics[width=\hsize]{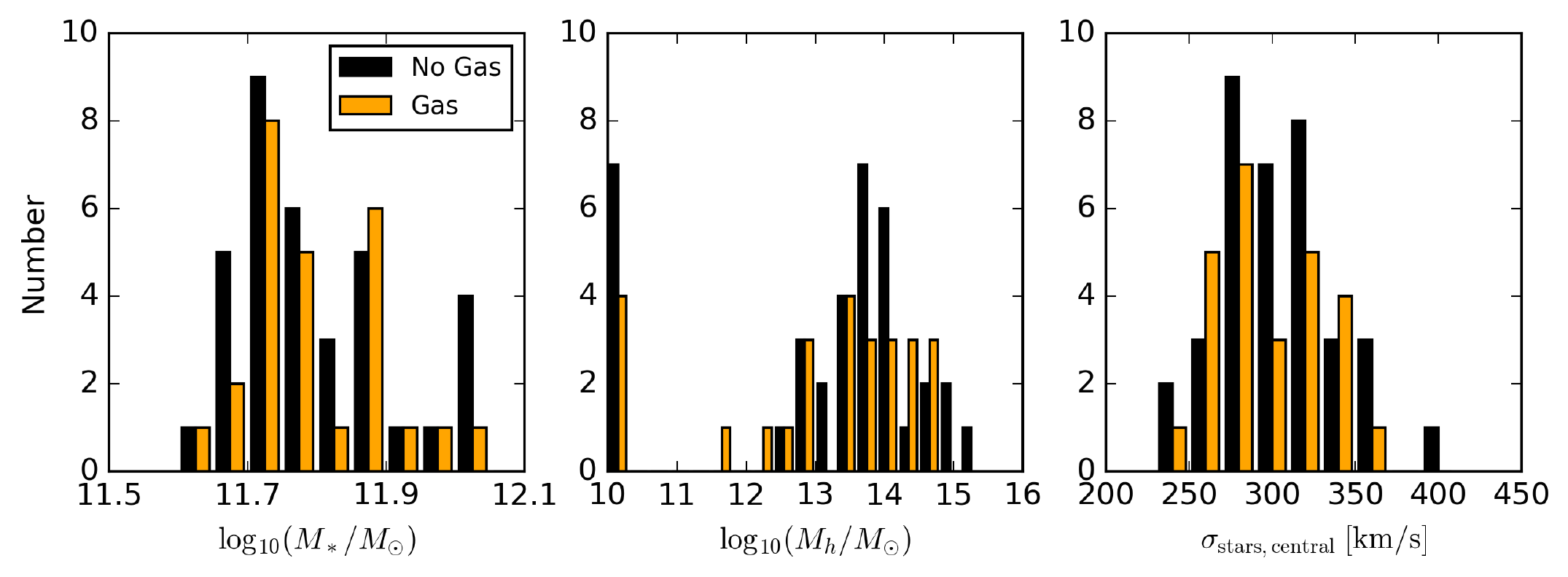}
\end{center}
\caption{The distributions of stellar mass (left), halo mass (middle), and central stellar velocity dispersion (right) for the MASSIVE galaxies with (orange) and without (black) warm ionized gas detections. After requiring measurements of the three physical properties to be available, there are 26 galaxies with gas and 36 galaxies without gas that comprise the distributions shown. Non-parametric tests reveal that each set of two distributions is consistent with being drawn from the same underlying parent population (see text). Note that a halo mass of $10^{10}M_{\odot}$ is an arbitrary value for galaxies with no identified friends \citep{crook07,crook08}.}
\label{fig:stats}
\end{figure*}

We are clearly more likely to detect warm ionized gas in MASSIVE fast rotators than slow rotators. We now ask whether there are other defining characteristics of the galaxies with warm ionized gas since that may provide indirect clues about the origin of the gas. For example, if cooling flows represent an important gas origin scenario for MASSIVE galaxies, then we may expect different parent populations of halo masses \citep[e.g.,][]{keres05}. If instead the dominant factor is whether a galaxy can recycle and retain the mass lost by its old stellar population, then we might expect different underlying distributions of stellar velocity dispersion. We thus investigate the following physical properties: stellar mass, halo mass, and central stellar velocity dispersion, the distributions of which are shown in \autoref{fig:stats}. We used three different non-parametric tests (two-sample Kolmogorov-Smirnov, Mann-Whitney-Wilcoxon, and Anderson-Darling) to check for statistically significant differences between the distributions of the three physical properties for the MASSIVE sample with and without gas. All three tests returned p-values $>0.01$ in every case, which suggests no statistically significant differences between the underlying parent populations.

\begin{sidewaystable*}
\vspace{100mm}
%\begin{table*}
%\tiny 
%\begin{center}
\caption{Sample of MASSIVE Galaxies with Warm Ionized Gas\label{tab:sample}}
\resizebox{\linewidth}{!}{
\begin{tabular}{ccccccccccccccccc}\hline
Galaxy & $D$ & $R_e$ & $R_{\rm gas}$ & $M_K$ & $M_*$ & $M_h$ & $M_{\rm gas}$ & $\mathcal{L}_{\rm H\beta}$ & $\sigma_{\rm stars,central}$ & [\ion{O}{3}]/H$\beta$ & $\rm PA_{\rm gas}$ & $\rm PA_{\rm stars}$ & $\rm PA_{\rm gas-stars}$ & FR/SR & Extended & Rotating \\
$-$ & [Mpc] & [kpc] & [kpc] & [mag] & [$\log M_{\odot}$] & [$\log M_{\odot}$] & [$\log M_{\odot}$] & [$\log$ erg s$^{-1}$] & [km s$^{-1}$] & $-$  & [deg] & [deg] & [deg] & $-$ & $-$ & $-$\\
(1) & (2) & (3) & (4) & (5) & (6) & (7) & (8) & (9) & (10) & (11) & (12) & (13) & (14) & (15) & (16) & (17) \\\hline\hline
NGC0708 & 69.0 & 16.32 & 4.09 & -25.65 & 11.75 & 14.5 & 5.52 & 39.1 & 248.03 & 2.14 & 356.8$\pm$1.6 & 126.0$\pm$17.5 & 50.8$\pm$17.6 & SR & Yes & Yes \\
NGC1167 & 70.2 & 9.97 & 18.21 & -25.64 & 11.74 & 13.0 & 5.73 & 39.3 & 231.81 & 5.71 & 284.3$\pm$15.3 & 254.0$\pm$3.8 & 30.3$\pm$15.8 & FR & Yes & Yes \\
NGC1453 & 56.4 & 7.63 & 8.41 & -25.67 & 11.75 & 13.6 & 5.82 & 39.4 & 341.04 & 2.59 & 312.5$\pm$3.1 & 35.0$\pm$3.2 & 97.5$\pm$4.4 & FR & Yes & Yes \\
NGC2208 & 84.1 & 10.91 & 8.86 & -25.63 & 11.74 & 10.0 & 5.55 & 39.1 & 280.34 & 2.25 & 241.8$\pm$13.4 & 95.0$\pm$9.5$^{\dagger}$ & $-$ & SR & Yes & Yes \\
NGC2320 & 89.4 & 8.16 & 5.65 & -25.93 & 11.87 & 14.1 & 5.57 & 39.1 & 345.86 & 2.66 & 148.2$\pm$16.5 & 147.0$\pm$4.0 & 1.2$\pm$17.0 & FR & Yes & Yes \\
NGC2783 & 101.4 & 18.24 & 2.85 & -25.72 & 11.78 & 12.3 & 5.54 & 39.1 & 274.76 & $>$1.18 & 148.5$\pm$5.0 & 269.5$\pm$11.0 & 59.0$\pm$12.1 & SR & Yes & Yes \\
NGC7426 & 80.0 & 7.26 & 5.74 & -25.74 & 11.79 & 13.7 & 5.71 & 39.3 & 306.75 & 2.66 & 261.7$\pm$3.1 & 251.0$\pm$1.2 & 10.7$\pm$3.3 & FR & Yes & Yes \\
NGC7550 & 72.7 & 9.64 & 2.35 & -25.43 & 11.65 & 11.8 & 5.32 & 38.9 & 270.73 & 3.31 & 238.7$\pm$10.4 & 155.2$\pm$15.5$^{\dagger}$ & $-$ & SR & Yes & Yes \\
NGC1700 & 54.4 & 6.06 & 12.27 & -25.60 & 11.72 & 12.6 & 4.82 & 38.4 & 262.74 & $>$0.63 & $-$ & 268.0$\pm$1.8 & $-$ & FR & Yes & No \\
NGC6482 & 61.4 & 4.44 & 16.46 & -25.60 & 11.72 & 13.0 & 5.61 & 39.2 & 330.44 & 1.26 & $-$ & 63.5$\pm$2.0 & $-$ & SR & Yes & No \\
NGC7265 & 82.8 & 10.93 & 7.41 & -25.93 & 11.87 & 14.6 & 5.08 & 38.6 & 256.21 & $>$1.20 & $-$ & 259.5$\pm$9.5 & $-$ & SR & Yes & No \\
NGC0315 & 70.3 & 8.42 & 1.68 & -26.30 & 12.03 & 13.4 & 4.74 & 38.3 & 353.32 & $>$1.06 & $-$ & 222.0$\pm$7.2 & $-$ & SR & No & No \\
NGC0383 & 71.3 & 7.09 & 1.73 & -25.81 & 11.82 & 14.3 & 4.60 & 38.1 & 304.02 & $>$1.56 & $-$ & 140.5$\pm$2.8 & $-$ & FR & No & No \\
NGC0410 & 71.3 & 11.30 & 1.79 & -25.90 & 11.86 & 14.3 & 4.66 & 38.2 & 313.07 & $>$0.97 & $-$ & 161.0$\pm$18.5 & $-$ & SR & No & No \\
NGC0533 & 77.9 & 15.29 & 1.88 & -26.05 & 11.92 & 13.8 & 4.96 & 38.5 & 304.81 & 1.44 & $-$ & 51.2$\pm$5.1$^{\dagger}$ & $-$ & SR & No & No \\
NGC0547 & 74.0 & 7.13 & 1.81 & -25.83 & 11.83 & $-$ & 4.58 & 38.1 & 277.65 & $>$2.14 & $-$ & 176.5$\pm$21.8 & $-$ & SR & No & No \\
NGC1060 & 67.4 & 12.98 & 1.76 & -26.00 & 11.90 & 14.0 & 4.19 & 37.7 & 325.91 & 0.88 & $-$ & 74.0$\pm$7.4$^{\dagger}$ & $-$ & SR & No & No \\
NGC3805 & 99.4 & 7.34 & 2.22 & -25.69 & 11.76 & 14.7 & 4.58 & 38.1 & 280.80 & 1.25 & $-$ & 61.5$\pm$1.5 & $-$ & FR & No & No \\
NGC3842 & 99.4 & 10.20 & 2.11 & -25.91 & 11.86 & 14.7 & 5.38 & 38.9 & 272.95 & 1.50 & $-$ & 1.6$\pm$0.2$^{\dagger}$ & $-$ & SR & No & No \\
NGC5129 & 107.5 & 10.11 & 2.32 & -25.92 & 11.86 & 10.0 & 4.66 & 38.2 & 277.81 & $>$1.21 & $-$ & 3.0$\pm$2.5 & $-$ & FR & No & No \\
NGC5208 & 105.0 & 5.41 & 1.48 & -25.61 & 11.73 & 13.0 & 4.27 & 37.8 & 274.84 & 2.09 & $-$ & 343.0$\pm$1.5 & $-$ & FR & No & No \\
NGC5322 & 34.2 & 2.42 & 0.60 & -25.51 & 11.68 & 13.5 & 4.08 & 37.6 & 258.94 & $>$1.13 & $-$ & 267.5$\pm$7.8 & $-$ & SR & No & No \\
NGC6223 & 86.7 & 7.85 & 2.05 & -25.59 & 11.72 & 13.5 & 5.14 & 38.7 & 294.97 & $>$0.72 & $-$ & 288.0$\pm$2.2 & $-$ & FR & No & No \\
NGC6375 & 95.8 & 9.08 & 2.26 & -25.53 & 11.69 & 10.0 & 4.76 & 38.3 & 251.26 & $>$2.99 & $-$ & 322.0$\pm$7.0 & $-$ & FR & No & No \\
NGC7052 & 69.3 & 8.98 & 1.65 & -25.67 & 11.75 & 10.0 & 3.63 & 37.2 & 314.09 & $>$1.26 & $-$ & 61.5$\pm$2.2 & $-$ & SR & No & No \\
NGC7436 & 106.6 & 12.39 & 2.48 & -26.16 & 11.97 & 14.2 & 4.59 & 38.1 & 311.13 & $>$1.07 & $-$ & 197.5$\pm$10.2 & $-$ & SR & No & No \\
NGC7626 & 54.0 & 6.21 & 1.16 & -25.65 & 11.75 & 13.9 & 3.52 & 37.1 & 305.02 & $>$0.94 & $-$ & 10.5$\pm$1.1$^{\dagger}$ & $-$ & SR & No & No \\
UGC02261 & 70.6 & 5.32 & 1.68 & -25.61 & 11.73 & 11.8 & 4.09 & 37.6 & 308.33 & $>$0.65 & $-$ & 66.0$\pm$2.5 & $-$ & FR & No & No \\
\end{tabular}}
\tablecomments{Properties of the MASSIVE sample with warm ionized gas. (1) Galaxy name. (2) Distance taken from \citet{ma14} [Mpc]. (3) Stellar effective radius [kpc]. (4) Maximum gas radius [kpc]. (5) Absolute $K$-band magnitude [AB mag]. (6) Logarithm of stellar mass. (7) Logarithm of the dark matter halo mass. (8) Logarithm of the warm ionized gas mass, which is derived using the H$\beta$ luminosity according to \autoref{eq:mgas}. (9) Logarithm of H$\beta$ luminosity. (10) Central stellar velocity dispersion [km s$^{-1}$]. (11) [\ion{O}{3}]/H$\beta$ flux ratio in the central spectrum (lower limit if H$\beta$ is non-detected). (12) Kinematic position angle of rotating warm ionized gas measured counterclockwise from north to the redshifted emission side [deg]. (13) Kinematic position angle of the stars measured counterclockwise from north to the redshifted emission side [deg]. (14) Kinematic misalignment angle of the warm ionized gas relative to the stars [deg]. (15) Kinematic classification as fast or slow rotator from \citet{veale16} based on angular momentum within $R_e$. (16) Whether the warm ionized gas is extended beyond the central three dithered fibers. (17) Whether the warm ionized gas shows regular rotation. In column (12), a $^{\dagger}$ identifies ``non-rotators" for which \texttt{kinemetry} was unable to determine a kinematic stellar position angle; for these galaxies, their photometric stellar position angles are given instead with an assumed 10\% fractional uncertainty (see \autoref{sec:kin} for details).}
\label{tab:det}
%\end{center}
%\end{table*}
\end{sidewaystable*}

\subsection{Spatial Extent}\label{sec:extent}
Because of the roughly $2\times2$ arcmin$^2$ field of view of the Mitchell Spectrograph, we are able to study the spatial extent of the gas out to generally $2-4\times R_e$. We use the binned IFU datasets described in \autoref{sec:sample} to determine via visual inspection the radially furthest bin in which an [\ion{O}{2}] emission line exists. 

Of the 28 MASSIVE galaxies with warm ionized gas, 11 galaxies have spatially extended gas and the remaining 17 galaxies have [\ion{O}{2}] emission only in a few fibers near their photometric centers. For the 11 spatially extended galaxies, we adopted the midpoint of the furthest bin with reliable [\ion{O}{2}] emission as the maximum radial extent of the warm ionized gas ($R_{\rm gas}$), and for the 17 compact cases, we assumed an upper limit on $R_{\rm gas}$ of 5 arcsec (recall that the fiber radii are 2 arcsec, but there are usually three dithers). The adopted $R_{\rm gas}$ values are given in \autoref{tab:det}. The maximum physical extents span a wide range from only 0.6 kpc up to 18.2 kpc, with a median of 2.3 kpc; likewise, the $R_{\rm gas}/R_e$ fractions run from 0.12 up to 3.71, with a median of 0.25. 

The furthest radial bin with detectable [\ion{O}{2}] emission contains on the order of $\sim100$ stacked individual fibers. Since we are binning so many fibers, a few of which can have detected emission (occasionally false positives), we ask whether the individual fiber detections might dominate the signal detected in the binned spectrum. In fact, the individual fibers constitute $<10\%$ of the integrated flux of the line detected in the binned spectrum. Thus, the binned detection suggests that there is extended low-surface-brightness emission filling a large fraction of the bin.

\subsection{Comparison to SAURON and ATLAS$^{\rm 3D}$}\label{sec:sauron}
Our $\sim39\pm6\%$ global detection fraction is lower than the $36/48=75\pm6\%$ detection fraction for SAURON's less massive early-type galaxies \citep{sarzi06}. While the \citet{sarzi06} work has not been extended to the full ATLAS$^{\rm 3D}$ sample, \citet{davis11} report that $67\pm3\%$ of fast rotators and $86\pm6\%$ of slow rotators in ATLAS$^{\rm 3D}$ have detectable warm ionized gas emission. Similarly, \citet{nyland16} report a warm ionized gas detection fraction of $54/101\approx53\pm5\%$ for the subset of ATLAS$^{\rm 3D}$ galaxies with radio observations. Both of these quoted ATLAS$^{\rm 3D}$ detection fractions are higher than ours. In this section, we will explore the detectability and resolvability of ATLAS$^{\rm 3D}$ and SAURON gas reservoirs if they instead existed in the more luminous and distant MASSIVE galaxies.\footnote{The median distances of the SAURON and ATLAS$^{\rm 3D}$ samples are 16 Mpc and 24 Mpc, respectively \citep{sarzi06,cappellari11}, substantially lower than the median distance of 80 Mpc for the MASSIVE sample \citep{ma14}.} We compare with an ATLAS$^{\rm 3D}$ catalog of ionized gas measurements (M. Sarzi, private communication). We only consider ATLAS$^{\rm 3D}$ galaxies that have detected warm ionized gas by requiring that the spatially integrated [\ion{O}{3}] EW $>0.02$\AA\ as in \citet{davis11}; there are 183 such objects.

First, if the ATLAS$^{\rm 3D}$ gas reservoirs were relocated to 80 Mpc (the median distance of the MASSIVE sample), would they be spatially resolved with our observations? Consider first that the 183 ATLAS$^{\rm 3D}$ galaxies with detected warm ionized gas span a range of physical radii from $0.2-4.4$ kpc, with a median of $1.6$ kpc. With a few exceptions on the smaller end, these values are consistent with the physical radii of MASSIVE compact (unresolved) gas reservoirs, whose upper limits on $R_{\rm gas}$ of five arcsec correspond to physical radii of $0.6-2.5$ kpc, with a median of $1.8$ kpc. Indeed, we find that $121/183\approx66\%$ of ATLAS$^{\rm 3D}$ gas reservoirs would have a radial gas extent of less than five arcsec if they were relocated to 80 Mpc, which means that the majority of the ATLAS$^{\rm 3D}$ sample would be classified as spatially unresolved in our observations.

Second, would our observations be sensitive enough to detect emission from ATLAS$^{\rm 3D}$ gas reservoirs, if they were relocated to the median MASSIVE distance of 80 Mpc? Only $101/183\approx55\%$ of ATLAS$^{\rm 3D}$ gas reservoirs shifted to 80 Mpc would have [\ion{O}{3}] EW above 0.5\AA\ within a single Mitchell fiber. If we crudely convert the ATLAS$^{\rm 3D}$ [\ion{O}{3}] EW to [\ion{O}{2}] EW by multiplying the former by a factor of six, then $123/183\approx67\%$ of ATLAS$^{\rm 3D}$ gas reservoirs relocated to 80 Mpc have [\ion{O}{2}] EW above 2\AA\ within a single Mitchell fiber.\footnote{The factor of six assumes [\ion{O}{2}]/[\ion{O}{3}]$=2$, which is typical for our MASSIVE galaxies as well as stacked SDSS spectra of extended LINER-like objects \citep[without reddening corrections;][]{johansson16}. We also assume that the stellar continuum is a factor of three fainter near [\ion{O}{2}] as compared to near [\ion{O}{3}] (see \autoref{sec:sens}).} This exercise suggests that the approximately factor of two lower warm gas detection fraction for MASSIVE galaxies, as compared to both the SAURON and ATLAS$^{\rm 3D}$ samples, might simply be attributable to sensitivity differences. Therefore, while the warm ionized gas detection fraction might indeed drop as a function of mass, we do not have good enough statistics to robustly constrain that expected trend.

We also note that the warm ionized gas detection fractions of fast rotators and slow rotators in ATLAS$^{\rm 3D}$ are comparable, which is not the case for MASSIVE. In addition, the slow rotator detection fraction is much higher for ATLAS$^{\rm 3D}$ ($\sim86\%$) than for MASSIVE ($\sim28\%$). These differences between the two surveys appear to be significant even though there are relatively few fast rotators in MASSIVE and relatively few slow rotators in ATLAS$^{\rm 3D}$. While the distributions of [\ion{O}{3}] EW for slow and fast rotators in ATLAS$^{\rm 3D}$ are statistically indistinguishable (according to three non-parametric tests), there is a longer tail toward high EWs for the fast rotators. Since we are not as sensitive, it is possible that a large fraction of ATLAS$^{\rm 3D}$ slow rotators with relatively low [\ion{O}{3}] EW would not be detected by us. Alternatively, it may be that the slow rotators have patchier gas distributions than the fast rotators, and thus our larger spaxels hurt us disproportionately.

\subsection{Kinematics: Gas versus Stars}\label{sec:kin}
A comparison of the kinematics of the stars and the warm ionized gas can reveal clues about the origin of the gas. For this reason, we run the \texttt{fit\_kinematic\_pa} routine described in \citet{krajnovic06} to determine the kinematic position angle (PA) of the warm ionized gas for the eight galaxies in which the gas is rotating and extended.\footnote{The \texttt{fit\_kinematic\_pa} Python routine is publicly available at \url{http://www.ascl.net/1601.016}.} We define the kinematic gas PA as the angle between north and the line that bisects the gas velocity field, measured counterclockwise from north to the redshifted emission side. In order to alleviate the impact of false positive fibers at large galactocentric radius and explore how well the PA can be constrained, we run \texttt{fit\_kinematic\_pa} three times, applying different fiber selection criteria each time and then taking the mean and standard deviation of the resulting values. The resulting kinematic gas PAs are given in \autoref{tab:det}. Note that it is difficult to constrain the kinematic position angle of the gas in NGC 1167 because rotation is observed in the gas only out to $\sim1R_{\rm e}$ and because NGC 1167 harbors a nearly face-on disk (see also \autoref{sec:special}). 

We have also measured kinematic stellar PAs for MASSIVE galaxies using \texttt{kinemetry} \citep{krajnovic06}. These values are listed in \autoref{tab:det}; additional details and measurements for the full MASSIVE sample will be presented in Ene et al. (in preparation). To ensure consistency with the kinematic gas PAs, our kinematic stellar PAs are also measured counterclockwise from north to the redshifted emission side. MASSIVE is dominated by slow rotators, and in some cases, there simply is not a preferred axis of rotation. Of the four slow rotators with extended and rotating warm ionized gas, NGC 2208 and NGC 7550 do not have measurable kinematic stellar PAs and we therefore do not consider them further in this section. The kinematic and photometric stellar PAs of the two other slow rotators, NGC 0708 and NGC 2783, are misaligned. In contrast, all four of the MASSIVE fast rotators that we consider here (NGC 1167, NGC 1453, NGC 2320, and NGC 7426) have measurable stellar kinematic PAs, and their kinematic and photometric stellar PA values are well-aligned. 

\begin{figure} 
\begin{center}
\includegraphics[width=\hsize]{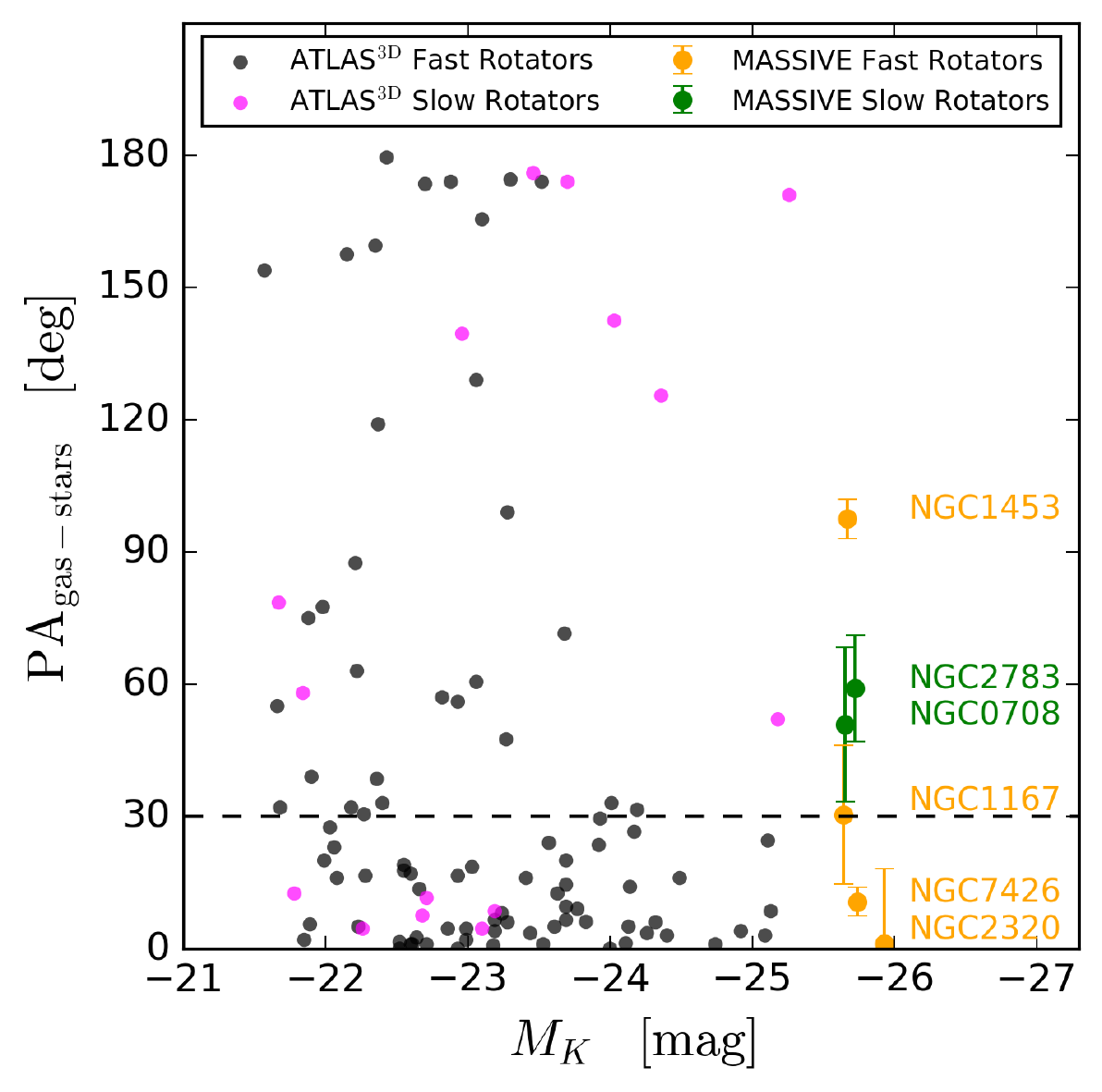}
\end{center}
\caption{Kinematic misalignment angle between the stars and spatially extended warm ionized gas reservoirs. MASSIVE slow rotators are shown as green circles and MASSIVE fast rotators as orange circles. ATLAS$^{\rm 3D}$ fast rotators (black circles) and slow rotators (magenta circles) are taken from \citet{davis11} for comparison purposes. The error bars for MASSIVE galaxies are a quadrature sum of the gas and stellar PAs. As with ATLAS$^{\rm 3D}$ fast rotators, MASSIVE fast rotators tend to have the warm ionized gas and stars aligned. The dashed horizontal line at 30 deg is the cutoff angle adopted from \citet{davis11} for stellar and gas alignment.}
\label{fig:kinematic}
\end{figure}

In \autoref{fig:kinematic}, we show the kinematic misalignment angle between the stars and warm ionized gas for MASSIVE slow and fast rotators as compared to ATLAS$^{\rm 3D}$ slow and fast rotators. Kinematic classifications of MASSIVE galaxies into slow and fast rotators are presented in \citet{veale16} whereas the kinematic misalignment angles and kinematic classifications of ATLAS$^{\rm 3D}$ galaxies are taken from \citet{krajnovic11}, \citet{davis11} and \citet{emsellem11}. Misalignment angles for MASSIVE galaxies are defined as $\rm PA_{\rm gas-stars}\equiv\rm PA_{\rm gas} - \rm PA_{\rm stars}$, where $\rm PA_{\rm gas}$ is the kinematic position angle of the warm ionized gas and $\rm PA_{\rm stars}$ is the position angle of the stars (kinematic when available, else photometric). The misalignment angles for all ATLAS$^{\rm 3D}$ galaxies are based on kinematic stellar PAs. All misalignment angles are defined to lie between $0$ and $180$ degrees. 

In general, misalignment between stars and gas points towards an external origin of the gas, while alignment between gas disks and stars suggests that the gas originated from the stars. However, see \citet{serra14} and \citet{lagos15} for the expected misalignment when gas cools out of the halo, and \citet{vandevoort15} for the timescales on which externally accreted gas is expected to become aligned with the stars due to gravitational torques. The MASSIVE fast rotators are predominantly aligned ($3/4 = 75 \pm 22\%$), which is consistent with the results from ATLAS$^{\rm 3D}$ that more massive fast rotators tend to host the warm ionized gas kinematically aligned with the stars \citep{davis11}. In contrast, the two MASSIVE slow rotators are consistent with misalignment between the stars and gas, which is in agreement with the interpretation that slow rotators predominantly get their gas via external means \citep[][]{sarzi06,davis11}. Our total number of rotating and spatially extended gas reservoirs is too small to say anything definitive about the distribution of misalignment angles. Nevertheless, the MASSIVE fast rotator NGC 1453 is especially interesting because its gas and stellar PAs are nearly perpendicular to each other; we will return to this galaxy in \autoref{sec:discussion}.

\subsection{Line Ratios and Radial Profiles}\label{sec:lineratios}
We now turn to the sources of ionization for the warm ionized gas in MASSIVE galaxies. There are five ionizing sources typically considered \citep[see][and references therein]{sarzi10}: photoionization by an active galactic nucleus \citep[AGN; e.g.,][]{ho97,ho08}, photoionization by evolved stars \citep[e.g.,][]{binette94,stasinska08}, photoionization by a young stellar component \citep[e.g.,][]{graves07,kaviraj07}, excitation by shocks \citep[e.g.,][]{koski76,ds95,allen08}, and thermal conduction via the hot X-ray-emitting gas \citep[e.g.,][]{sparks89,mathews03,sarzi13}. 

Since we do not have access to standard excitation diagrams that make use of the [\ion{N}{2}]$\lambda\lambda6548,6584$/H$\alpha$ ratio \citep[e.g.,][]{bpt81,veilleuxosterbrock87}, it is difficult to distinguish between the first four sources of ionization for our sample. Nevertheless, we can speculate using two alternative methods: (1) use non-standard excitation diagrams to find extraordinary cases where a single source of ionization can be unambiguously identified \citep[e.g., see][]{sarzi06,sarzi10,annibali10,yan12}, and (2) check whether the radial profile of the ionized gas flux follows the radial profile of the stellar continuum flux as probed by the emission line EW. As we obtain more X-ray observations of MASSIVE galaxies using \textit{Chandra} (PI: Goulding), we will be in a position to also examine trends between X-ray properties and the presence of warm ionized gas. 

The first clue about the sources of ionization comes from emission line ratios.\footnote{Due to the limited wavelength coverage of our spectra, we can only measure the [\ion{O}{3}]/H$\beta$ and [\ion{N}{1}]/H$\beta$ ratios.} Since we can only compute robust line ratios for the central parts of our galaxies, we will use only the value in the central fiber for each galaxy. However, we should note that our 4 arcsec diameter central fiber typically covers the inner $\sim1.5$ kpc of our galaxies. In such a large physical aperture, even in cases where an AGN plays some role in the photoionization of the gas, there are likely to be contributions from other sources \citep[e.g.,][]{eracleous10}. Therefore, aside from relatively luminous Seyfert galaxies such as NGC 1167 (see below), we cannot say much about optical AGN activity from these data. In standard excitation diagrams \citep[e.g.,][]{bpt81,veilleuxosterbrock87,kewley06}, the [\ion{O}{3}]/H$\beta$ ratio is combined with another strong line, typically [\ion{N}{2}]/H$\alpha$, to distinguish between \ion{H}{2} regions, shocks, and an AGN. Star-forming galaxies follow a path toward higher [\ion{O}{3}]/H$\beta$ and lower [\ion{N}{2}]/H$\alpha$ values as metallicity decreases and the ionization parameter increases \citep[e.g., see][]{kennicutt98,kewley02,kauffmann03,tremonti04,kewley06,moustakas06}. If we assume that the high stellar metallicity that characterizes our galaxies \citep{greene15} also applies to the gas, then we can more easily map the [\ion{O}{3}]/H$\beta$ ratio to sources of ionization \citep[see discussion in Section 6.2 of][]{nyland16}. High values of [\ion{O}{3}]/H$\beta$ ($\gtrsim3$) are likely to point towards AGN activity, while very low values ($\lesssim0.5$) would point towards star formation. 

In our sample, NGC 1167 has [\ion{O}{3}]/H$\beta\approx6$, which is indicative of AGN activity \citep{groves04a,groves04b}. The remaining galaxies have intermediate [\ion{O}{3}]/H$\beta$ measurements (or lower limits when H$\beta$ is not detected). These intermediate [\ion{O}{3}]/H$\beta$ values are a defining feature of low-ionization nuclear emission regions \citep[LINERs; e.g., see][]{heckman80,kewley02}, which may arise from excitation by evolved stars, shocks, or a low-luminosity AGN \citep[e.g.,][]{leitherer99,dopita00,ho08,allen08,eracleous10,groves10}. In several of our galaxies, non-central fibers tend to also exhibit these so-called ``extended LINER-like" excitation ratios \citep[i.e., low ionization emission regions or LIERs;][]{belfiore16,johansson16}. In principle, the [\ion{N}{1}]/H$\beta$ ratio can help us distinguish between excitation grids for fast shocks and \ion{H}{2} regions \citep[see detailed discussion by][]{sarzi10}, but the [\ion{N}{1}] line is detected in the centers of only three galaxies. One of those galaxies is the aforementioned NGC 1167, which is already consistent with central excitation by an AGN based on its high [\ion{O}{3}]/H$\beta$ ratio alone. For the remaining two galaxies (NGC 0708 and NGC 1453), the combination of [\ion{O}{3}]/H$\beta$ and [\ion{N}{1}]/H$\beta$ ratios is more consistent with excitation grids for fast shocks than \ion{H}{2} regions, which again broadly suggests LINER-like emission.

\begin{figure*} 
\begin{center}
\includegraphics[width=0.9\hsize]{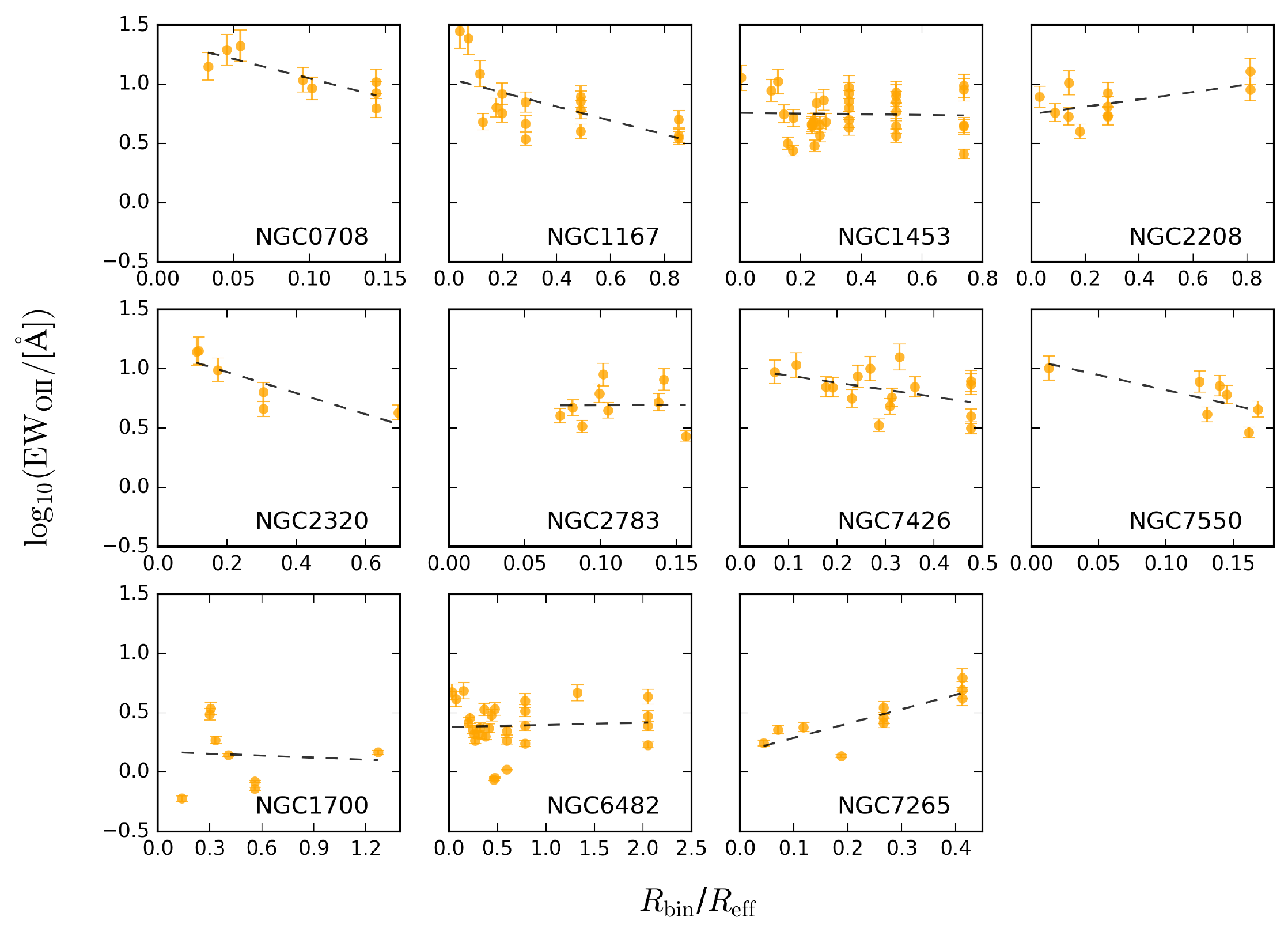}
\end{center}
\caption{Logarithmic radial profiles of the [\ion{O}{2}] EW for the 11 MASSIVE galaxies with extended warm ionized gas. The solid lines represent linear fits to the logarithmic radial profiles whereas the orange scatter points show the actual measurements made using the binned spectra. The linear model does not fully capture the complexity of the data in most cases, and a significant number of galaxies reveal non-flat radial profiles of the [\ion{O}{2}] equivalent width (with NGC 1167 clearly showing a non-linear rise toward the very center). Note that since $R_e$ and $R_{\rm gas}$ are different for every galaxy, the horizontal axis scale is also different in each subpanel.}
\label{fig:radialew}
\end{figure*}

Given the limitations of the line ratio approach for our dataset, we now turn to the spatial regime. If the source of ionization was centrally concentrated, then we would expect an upturn in emission line EW toward the center of the galaxy along with central excitation ratios suggestive of, say, an AGN. In the simplest case where a uniformly distributed old stellar population is responsible for ionizing the gas, we would expect a roughly constant radial EW profile. In \autoref{fig:radialew}, we show the radial profiles of logarithmic [\ion{O}{2}] EW as well as the best-fit straight lines for the 11 MASSIVE galaxies with extended warm ionized gas. Note that the radial scale of the EW profile is different for every galaxy because both $R_e$ and $R_{\rm gas}$ are different. A straight line model for $\log_{10}$(EW$_{\rm OII}$/[\AA]) as a function of galactocentric radius does not always capture the complexity of the data, but at least half of the cases shown are consistent with a slope of zero.

In several cases, the data unambiguously reveal a non-constant trend in the equivalent width radial profile. An example is NGC 1167 in which the [\ion{O}{2}] EW rises non-linearly and rapidly toward the center of the galaxy, suggesting that an AGN dominates as the source of ionizing photons near the center of this galaxy \citep[this upturn toward the center is also seen in the radial EW profile of H$\alpha$ by][in the CALIFA survey]{gomes16}. A similar phenomenon can be seen for NGC 6482, but it is less clear due to the different radial scale of the EW profile. In contrast, NGC 7265 presents an increasing [\ion{O}{2}] equivalent width radial profile. Since the observed EW depends on both the surface brightness of the starlight and the absorption of the photoionizing continuum by the gas (which itself depends on the geometry of the gas, dust, etc.), the expected radial dependence of the EW in different heating scenarios is not obvious \citep[e.g.,][]{yan12,papaderos13,gomes16,gomes16b}. In a simple geometric model in which the stars are spherically distributed and the gas is in a disk, the EW would actually rise with radius \citep{sarzi10}, but generally in the SAURON galaxies the EWs are flat or fall slightly with radius. 

Taking post-asymptotic giant branch (pAGB) stars as the main source of photoionization, stellar population synthesis models predict an H$\alpha$ EW range of $\sim 0.5-3$\AA\ \citep[e.g.,][]{gomes16,belfiore16}. Indeed, the stacked SDSS spectrum of extended LINER-like objects presented by \citet{johansson16}, which are presumably powered by pAGB stars, has an H$\alpha$ EW of 1.3\AA. Their empirical spectrum also has an H$\alpha$/[\ion{O}{2}] EW ratio of $\approx0.3$, which corresponds to a predicted [\ion{O}{2}] EW of $\approx4.3$\AA\ (see their Table 2 and Figure 3). This predicted [\ion{O}{2}] EW, which is generally attributed solely to pAGB stars, is well above our own [\ion{O}{2}] EW detection limit of 2\AA. Thus, there is apparently a real difference between the emission line EWs in MASSIVE galaxies and the average galaxy considered by \citet{johansson16}. There are many possible sources for this discrepancy: the typical line EW due to pAGB stars might be different in the most massive galaxies (e.g., because of a higher metal content or different mean stellar ages), the total amount of warm ISM available to be ionized might genuinely be lower, or the line ratios could simply be different as a function of galaxy mass (again due to differences in metallicity or the ionizing spectrum from pAGB stars).

\subsection{Warm Ionized Gas Masses}
For all 28 MASSIVE galaxies that host warm ionized gas, we estimate gas masses as follows. Within $R_{\rm gas}$, we take all fibers which have $\mathcal{S}\geq3$ for [\ion{O}{2}] and assume that [\ion{O}{2}]/H$\beta=6$ (based on the mean of the central fibers of NGC 1453, in which we have the greatest number of H$\beta$ detections) to compute an H$\beta$ luminosity. To compute the total warm ionized gas mass as traced by the H$\beta$ recombination line, we use equation (1) of \citet{nesvadba11}, which we rewrite here for completeness:
\begin{equation}\label{eq:mgas}
\mathcal{M}_{\rm H\beta} = 28.2 \times 10^8\; \mathcal{L}_{\rm H\beta,43} \;n_{\rm e,100}^{-1} \;M_{\odot}\;.
\end{equation}
Here, $\mathcal{L}_{\rm H\beta,43}$ is the H$\beta$ luminosity in units of $10^{43}$ erg s$^{-1}$ and $n_{\rm e,100}$ is the electron number density in units of 100 cm$^{-3}$. Due to the limited wavelength coverage of our spectra, we do not have independent constraints on $n_e$ for each galaxy, so we adopt the common value of $n_e=100$ cm$^{-3}$ based on observations of other early-type galaxies \citep[see][]{mathews03}. The derived warm ionized gas masses and corresponding $\mathcal{L}_{\rm H\beta}$ are given in \autoref{tab:det}. There is roughly an order of magnitude systematic uncertainty in the gas masses due to the unknown gas density and line ratios, but the derived masses are quite consistent at this level with previous work \citep[see also][]{goudfrooij99}.

\begin{figure*} 
\begin{center}
\includegraphics[width=0.9\hsize]{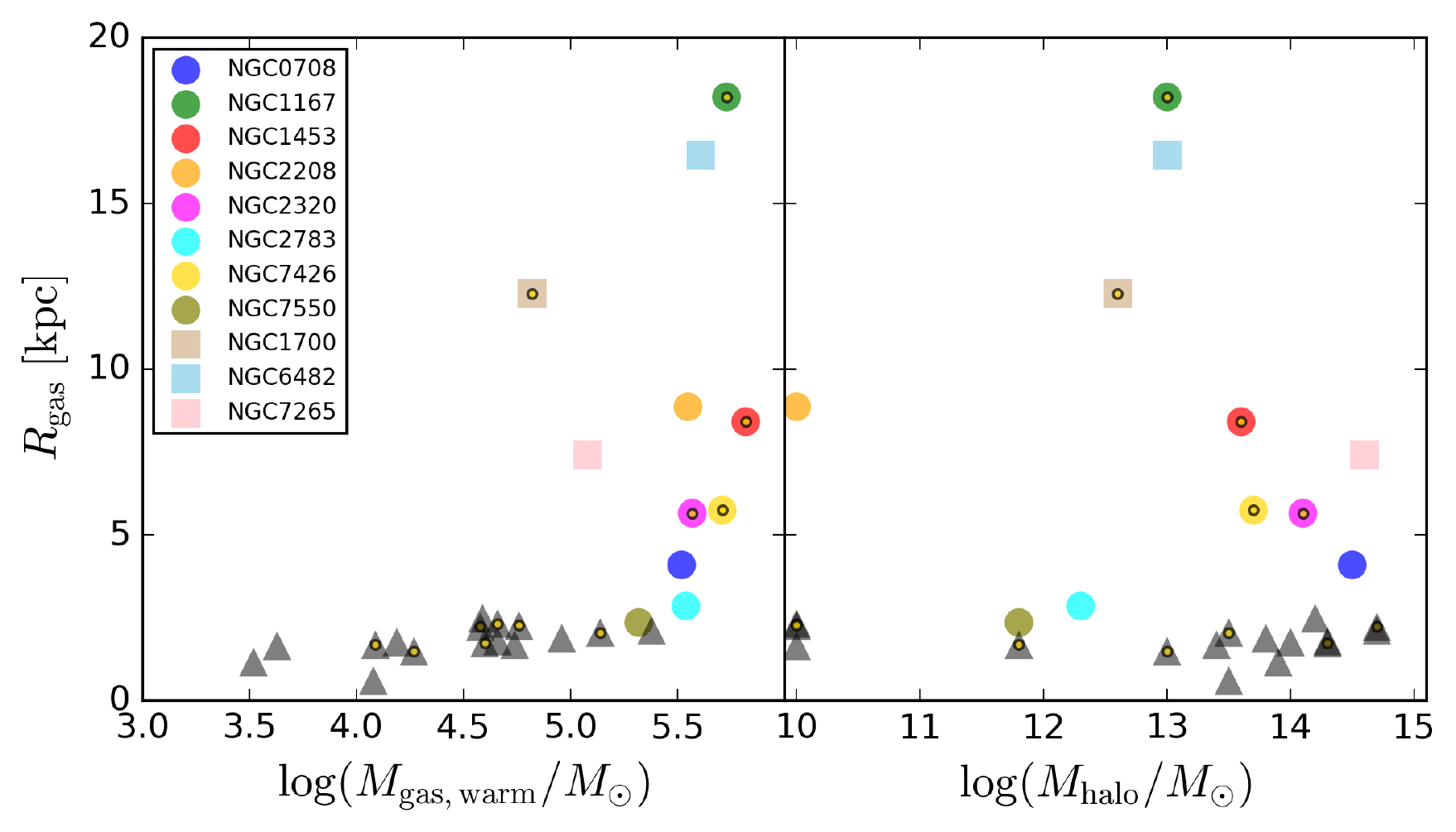}
\end{center}
\caption{Maximum radial extent of the gas versus warm ionized gas mass (left) and halo mass (right). Circles represent galaxies with rotating warm ionized gas, squares represent galaxies with non-rotating but extended gas reservoirs, and triangles are lower limits for galaxies hosting compact nuclear gas reservoirs. Little gold circles mark the fast rotators.}
\label{fig:mass}
\end{figure*}

\autoref{fig:mass} shows the maximum radial extent of the warm ionized gas as a function of the warm ionized gas mass and the halo mass. The warm ionized gas masses span $10^4-10^5M_{\odot}$. As expected, cases of compact centralized gas detections have lower warm ionized gas masses than spatially extended gas reservoirs. The warm ionized gas tends to have the greatest radial extent in galaxies with a patchy, non-rotating spatial gas distribution. 

\section{Discussion}\label{sec:discussion}
We will attempt to use the morphology, kinematics, and ionization of the gas to infer some constraints on its origins. We consider origins both internal (stellar mass loss or cooling from a hot halo) and external (merging or gas accretion). 

\subsection{Gas Morphology}
Among our sample of 28 MASSIVE galaxies with detected warm ionized gas, there exist a wide variety of gas morphologies. It is instructive to consider whether the morphology of the gas can help distinguish between various possible gas origin scenarios. Motivated by the work of \citet{macchetto96}, we can roughly classify all of our different gas reservoirs into three broad categories: (1) smooth, extended, and rotating disks, (2) extended and filamentary/patchy gas, and (3) unresolved/small reservoirs. The majority of our detections are of the unresolved/compact variety ($17/28\approx61\pm9\%$). In our spatially resolved cases, we are detecting the most extreme gas reservoirs in terms of spatial extent, and filamentary emission is less frequently detected ($3/28\approx11\pm6\%$) than smooth rotating emission ($8/28\approx29\pm9\%$). Our detection fractions for each of the three morphological categories are roughly in agreement with the similarly constructed categories of \citet{macchetto96}. 

There is little that we can say about the unresolved gas, except that it likely resides in the small disks that are seen in 75\% of early-type galaxies \citep[e.g., see ][]{vandokkum95,ferrarese99,martini03b}.  These dust disks can have semi-major axes up to several hundred parsecs \citep[see Table 2 of][]{vandokkum95}. Given that our 2 arcsec radius fibers correspond to $\sim0.75$ kpc at 80 Mpc and that we cannot resolve $R_{\rm gas}$ below $1-3\times0.75$ kpc (based on how many central fibers the gas is detected in), it is entirely possible that in our 18 galaxies with compact gas reservoirs, the ionized gas emission comes from such a central dust disk. It is also possible that there is extended but non-detected gas in these galaxies (see \autoref{sec:sauron}).

\subsection{Origin of the Gas}\label{sec:discorigin}
We will now step through the different possible gas origin scenarios by presenting emblematic examples of each.

\subsubsection{External accretion versus internal origins}
Misalignment angles are often used as clues to the origin of the gas. While alignment may point to internal or external processes, misalignment is a clear sign of external accretion \citep[e.g.,][]{sarzi06,davis11}. In contrast to external accretion, stellar mass loss is a prominent internal source of gas. It is difficult to say for sure which of our galaxies have gas reservoirs consistent with an internal recycling origin. Although kinematic alignment between the stars and gas is a clue in favor of stellar mass loss, such alignment can also occur in merger remnants within a few dynamical times due to gravitational torques exerted on the gas by the stars \citep[e.g.,][]{vandevoort15}. 

An extreme example of external accretion is the MASSIVE fast rotator NGC 1453 in which the warm ionized gas is kinematically misaligned with respect to the stars by $\sim98$ deg. \citet{buson93} and \citet{zeilinger96} also found that the warm ionized gas is misaligned relative to the stars in NGC 1453, though with a larger misalignment angle of $\approx128$ deg. It is unlikely that stellar mass loss from the old stellar population is the primary source of the warm ionized medium in NGC 1453 because such recycled gas would tend to remain in the potential well of the stars, leading to photometric and kinematic alignment between the stars and gas. 

\subsubsection{Cooling flows}
In NGC 6482, the gas is patchy and its velocity structure may hint at cooling flows as the dominant origin scenario \citep[e.g., see][]{fabian94}. NGC 6482 is known to be a fossil group system; it is a single elliptical galaxy with a group-like X-ray halo and in which all other group members are at least two optical magnitudes fainter \citep[see][]{mulchaey00,khosroshahi04}. This galaxy is also known to harbor a patchy dust reservoir \citep{alamomartinez12}. We do not have the spatial resolution to study the detailed morphology of the gas, but if the origin of the warm ionized gas is indeed via cooling flows, then we would expect to see a connection with the hot X-ray-emitting halo gas. For example, \citet{mcdonald11a} find a clear relationship between H$\alpha$ filaments and the X-ray properties of ``cool core" clusters \citep[negative radial temperature gradients; e.g., see][]{hudson10}. Future comparisons of deep narrow-band and X-ray imaging could help us test the cooling flow origin hypothesis in MASSIVE galaxies like NGC 6482. Also powerful would be comparisons to known ``cool-core" cluster galaxies in MASSIVE which have warm ionized gas detections (NGC 0315, NGC 0533, NGC 0547 and NGC 0708) using deep high-resolution images that clearly reveal the morphology of the gas.

\subsection{Special Cases}\label{sec:special}
The MASSIVE fast rotator NGC 1167 represents an interesting case of both gas rotation near the center of the galaxy and a patchy distribution of gas at larger galactocentric radius. The galaxy itself is likely a face-on disk, and exhibits extended UV emission in images taken with the \textit{Galaxy Evolution Explorer} \citep[\textit{GALEX};][]{martin05}. Interestingly, \citet{struve10} detected a $\sim160$ kpc disk of neutral hydrogen in NGC 1167 and suggested satellite accretion as its origin. In total, the warm ionized gas in NGC 1167 reaches roughly 18 kpc, but rotation is evident in the gas only within about 10 kpc. Since gas is detected in the outermost bins for NGC 1167 and there might be more detectable emission outside the field of view of the Mitchell Spectrograph, our $R_{\rm gas}$ can be considered a lower limit for this galaxy. We see evidence of an AGN based on the [\ion{O}{3}]/H$\beta$ line ratio, and our detection of [\ion{O}{2}] in binned fibers at large galactocentric radius hints at excitation by \ion{H}{2} regions. As part of the CALIFA survey, \citet{gomes16} found clear evidence of localized H$\alpha$ clumps at large galactocentric radii (see their Figure 1), which they attribute to low-level ongoing star formation. They also comment on the patchiness of the gas by classifying NGC 1167 as having a ``perturbed" rotational pattern (see their Table 1). This is one clear case where residual star formation may accompany and indeed ionize the gas \citep[an assertion further supported by the discovery of $3.3\times10^{8}M_{\odot}$ worth of molecular gas, and a mid-infrared SFR of $\sim0.3M_{\odot}$ yr $^{-1}$;][]{osullivan15,davis16a}.

Another very interesting case is NGC 7265. It is unclear what the source of the gas is in NGC 7265 since it is redshifted with respect to the systemic velocity and has a very asymmetric spatial extent with no blueshifted component. The gas velocities are consistent with the systemic velocity near the center of the galaxy but extend up to 250 km s$^{-1}$ at a radius of 7.4 kpc (see \autoref{fig:portraits}). It is tempting to think we are observing an outflow driven by an AGN \citep[e.g.,][]{ciotti16}, in analogy with the ATLAS$^{\rm 3D}$ galaxy NGC 1266 \citep{alatalo11,davis12,nyland13,alatalo15}. However, with only redshifted emission, we cannot rule out an inflow, rotation, or simply an ongoing merger. We do detect a compact radio source (Nyland et al. in preparation) and a soft nuclear X-ray source \citep{goulding16} in the center. Since we do not detect H$\beta$ we cannot say much about the nuclear line ratios. If this is an outflow, however, it is difficult to understand why we observe only redshifted emission.

\section{Future Work and Summary}\label{sec:summary}
We have investigated the existence and spatial distribution of warm ionized gas in a volume- and magnitude-limited sample of massive early-type galaxies with $\log(M_*/M_{\odot})>11.5$ ($M_K<-25.3$ mag) and distance $D<108$ Mpc using deep wide-field integral field spectroscopy. The main conclusions of this paper are:
\begin{enumerate}
\item We find that $28/74\approx38\pm6\%$ of MASSIVE early-type galaxies host warm ionized gas down to an [\ion{O}{2}] EW limit of $\sim2$\AA. In eight of these cases, the warm ionized gas exists in the form of a large rotating disk, whereas in three cases, the gas is extended but patchy/non-rotating. The MASSIVE galaxies NGC 1167, NGC 1700, and NGC 6482 host the most extended warm ionized gas reservoirs, out to roughly 18, 12 and 16 kpc, respectively. The remaining 17 galaxies host detectable warm ionized gas only near their centers. Our detection fraction is lower than the $\sim70\%$ measured by the SAURON and ATLAS$^{\rm 3D}$ surveys of nearer but less massive early-type galaxies. This difference is consistent with arising from our shallower surface brightness sensitivity. 
\item The warm ionized gas detection fraction does not appear to depend on stellar mass, halo mass, or central stellar velocity dispersion. However, there is a statistically significant ($3.7\sigma$) dependence in terms of net angular momentum: $\sim80\%$ of MASSIVE fast rotators host warm ionized gas compared to only $\sim30\%$ of MASSIVE slow rotators. The much lower detection fraction for MASSIVE slow rotators compared to ATLAS$^{\rm 3D}$ slow rotators ($\sim85\%$) can at least partially be attributed to sensitivity differences between the two surveys.
\item Of the four fast rotators with rotating and spatially extended gas, three show kinematic alignment between the stars and gas, which is consistent with previous findings by ATLAS$^{\rm 3D}$ that more massive fast rotators tend to have the gas and stars kinematically aligned. On the other hand, both slow rotators with measurable kinematic stellar and gas position angles show misalignment, which agrees with the idea that slow rotators generally get their gas through external means.
\item The most prominent gas excitation mechanisms are likely photoionization from evolved stars and fast shocks on extended kpc scales with possible additional contributions from low-luminosity AGN on nuclear scales. Using radial profiles of the [\ion{O}{2}] EW and non-standard excitation diagrams, we cannot easily pinpoint the sources of ionization for the gas. The one exception is NGC 1167, whose central line ratios and radial [\ion{O}{2}] EW profile suggest photoionization by an AGN for gas near the center. 
\item It is likely that a variety of physical processes are responsible for the origin and ionization of the gas in MASSIVE early-type galaxies. We see examples of gas with large misalignment angles that were likely acquired via external accretion, a $\sim$16 kpc-scale patchy gas reservoir likely cooling from a hot halo, and even a possible AGN-driven outflow.
\end{enumerate}

One exciting possibility opened up by our observations is to use the extended rotating disks we have uncovered to measure enclosed dynamical masses. These dynamical masses will be complementary to those that we are deriving from modeling the stars \citep[e.g.,][]{thomas16}, and will give us the opportunity to test the stellar dynamical models as well as examine the universality of the initial mass function \citep[e.g.,][]{treu10,cappellari12,conroy12}. Such dynamical modeling will also allow us to test whether what we are calling warm ionized gas disks are actually so-called ``red geysers," which are thought to be large-scale AGN-driven outflows \citep[][]{cheung16}.

Also of considerable interest is connecting our work on the warm ionized medium of MASSIVE early-type galaxies to other phases of their ISM. The relationship between the warm ionized and hot X-ray-emitting gas is still unclear \citep[e.g.,][]{forman85,heckman89,macchetto96,goudfrooij99,sarzi13,goulding16}, and we have approved \textit{Chandra} time (PI: Goulding) to increase the MASSIVE sample with hot gas measurements. Similarly, the occurrence rate of warm ionized gas (and dust) as a function of radio continuum detection may reveal relationships to accreting supermassive black holes and the presence of central dust disks; this will be explored in a future study. Finally, finding direct relationships between the warm ionized and cold gas phases (both molecular and atomic) is important for addressing the fate of the gas in these massive quiescent systems; our studies of CO emission from these galaxies probe fundamental questions along these lines \citep[][and Davis et al. in preparation]{davis16a}. 

\section*{Acknowledgements}
We thank Marc Sarzi for providing us with a catalog of ATLAS$^{\rm 3D}$ warm ionized gas measurements and for helpful suggestions. We also thank Duncan Forbes, Brent Groves, Yifei Jin, Davor Krajnovic and Freeke van de Voort for helpful comments. We thank the referee for a thorough and useful report that significantly improved the paper. VP thanks Bill Mathews, Joel Primack and X. Prochaska for helpful discussions. The MASSIVE survey is supported in part by NSF AST-1411945, NSF AST-1411642, HST-GO-14210, and HST-AR-14573.

\bibliographystyle{apj}
\bibliography{references}

\begin{thebibliography}{}
\expandafter\ifx\csname natexlab\endcsname\relax\def\natexlab#1{#1}\fi

\bibitem[{{Alamo-Mart{\'{\i}}nez} {et~al.}(2012){Alamo-Mart{\'{\i}}nez},
  {West}, {Blakeslee}, {Gonz{\'a}lez-L{\'o}pezlira}, {Jord{\'a}n}, {Gregg},
  {C{\^o}t{\'e}}, {Drinkwater}, \& {van den Bergh}}]{alamomartinez12}
{Alamo-Mart{\'{\i}}nez}, K.~A., {West}, M.~J., {Blakeslee}, J.~P., {et~al.}
  2012, \aap, 546, A15

\bibitem[{{Alatalo} {et~al.}(2011){Alatalo}, {Blitz}, {Young}, {Davis},
  {Bureau}, {Lopez}, {Cappellari}, {Scott}, {Shapiro}, {Crocker},
  {Mart{\'{\i}}n}, {Bois}, {Bournaud}, {Davies}, {de Zeeuw}, {Duc}, {Emsellem},
  {Falc{\'o}n-Barroso}, {Khochfar}, {Krajnovi{\'c}}, {Kuntschner}, {Lablanche},
  {McDermid}, {Morganti}, {Naab}, {Oosterloo}, {Sarzi}, {Serra}, \&
  {Weijmans}}]{alatalo11}
{Alatalo}, K., {Blitz}, L., {Young}, L.~M., {et~al.} 2011, \apj, 735, 88

\bibitem[{{Alatalo} {et~al.}(2013){Alatalo}, {Davis}, {Bureau}, {Young},
  {Blitz}, {Crocker}, {Bayet}, {Bois}, {Bournaud}, {Cappellari}, {Davies}, {de
  Zeeuw}, {Duc}, {Emsellem}, {Khochfar}, {Krajnovi{\'c}}, {Kuntschner},
  {Lablanche}, {Morganti}, {McDermid}, {Naab}, {Oosterloo}, {Sarzi}, {Scott},
  {Serra}, \& {Weijmans}}]{alatalo13}
{Alatalo}, K., {Davis}, T.~A., {Bureau}, M., {et~al.} 2013, \mnras, 432, 1796

\bibitem[{{Alatalo} {et~al.}(2015){Alatalo}, {Lacy}, {Lanz}, {Bitsakis},
  {Appleton}, {Nyland}, {Cales}, {Chang}, {Davis}, {de Zeeuw}, {Lonsdale},
  {Mart{\'{\i}}n}, {Meier}, \& {Ogle}}]{alatalo15}
{Alatalo}, K., {Lacy}, M., {Lanz}, L., {et~al.} 2015, \apj, 798, 31

\bibitem[{{Allen} {et~al.}(2008){Allen}, {Groves}, {Dopita}, {Sutherland}, \&
  {Kewley}}]{allen08}
{Allen}, M.~G., {Groves}, B.~A., {Dopita}, M.~A., {Sutherland}, R.~S., \&
  {Kewley}, L.~J. 2008, \apjs, 178, 20

\bibitem[{{Annibali} {et~al.}(2010){Annibali}, {Bressan}, {Rampazzo},
  {Zeilinger}, {Vega}, \& {Panuzzo}}]{annibali10}
{Annibali}, F., {Bressan}, A., {Rampazzo}, R., {et~al.} 2010, \aap, 519, A40

\bibitem[{{Athey} {et~al.}(2002){Athey}, {Bregman}, {Bregman}, {Temi}, \&
  {Sauvage}}]{athey02}
{Athey}, A., {Bregman}, J., {Bregman}, J., {Temi}, P., \& {Sauvage}, M. 2002,
  \apj, 571, 272

\bibitem[{{Bacon} {et~al.}(2001){Bacon}, {Copin}, {Monnet}, {Miller},
  {Allington-Smith}, {Bureau}, {Carollo}, {Davies}, {Emsellem}, {Kuntschner},
  {Peletier}, {Verolme}, \& {de Zeeuw}}]{bacon01}
{Bacon}, R., {Copin}, Y., {Monnet}, G., {et~al.} 2001, \mnras, 326, 23

\bibitem[{{Baldwin} {et~al.}(1981){Baldwin}, {Phillips}, \&
  {Terlevich}}]{bpt81}
{Baldwin}, J.~A., {Phillips}, M.~M., \& {Terlevich}, R. 1981, \pasp, 93, 5

\bibitem[{{Belfiore} {et~al.}(2016){Belfiore}, {Maiolino}, {Maraston},
  {Emsellem}, {Bershady}, {Masters}, {Yan}, {Bizyaev}, {Boquien}, {Brownstein},
  {Bundy}, {Drory}, {Heckman}, {Law}, {Roman-Lopes}, {Pan}, {Stanghellini},
  {Thomas}, {Weijmans}, \& {Westfall}}]{belfiore16}
{Belfiore}, F., {Maiolino}, R., {Maraston}, C., {et~al.} 2016, \mnras,
  arXiv:1605.07189

\bibitem[{{Binette} {et~al.}(1994){Binette}, {Magris}, {Stasi{\'n}ska}, \&
  {Bruzual}}]{binette94}
{Binette}, L., {Magris}, C.~G., {Stasi{\'n}ska}, G., \& {Bruzual}, A.~G. 1994,
  \aap, 292, 13

\bibitem[{{Bruzual} \& {Charlot}(2003)}]{bc03}
{Bruzual}, G., \& {Charlot}, S. 2003, \mnras, 344, 1000

\bibitem[{{Buson} {et~al.}(1993){Buson}, {Sadler}, {Zeilinger}, {Bertin},
  {Bertola}, {Danzinger}, {Dejonghe}, {Saglia}, \& {de Zeeuw}}]{buson93}
{Buson}, L.~M., {Sadler}, E.~M., {Zeilinger}, W.~W., {et~al.} 1993, \aap, 280,
  409

\bibitem[{{Caldwell}(1984)}]{caldwell84}
{Caldwell}, N. 1984, \pasp, 96, 287

\bibitem[{{Caon} {et~al.}(2000){Caon}, {Macchetto}, \& {Pastoriza}}]{caon00}
{Caon}, N., {Macchetto}, D., \& {Pastoriza}, M. 2000, \apjs, 127, 39

\bibitem[{{Cappellari}(2016)}]{cappellari16}
{Cappellari}, M. 2016, ArXiv e-prints, arXiv:1602.04267

\bibitem[{{Cappellari} \& {Emsellem}(2004)}]{cappellari04}
{Cappellari}, M., \& {Emsellem}, E. 2004, \pasp, 116, 138

\bibitem[{{Cappellari} {et~al.}(2011){Cappellari}, {Emsellem}, {Krajnovi{\'c}},
  {McDermid}, {Scott}, {Verdoes Kleijn}, {Young}, {Alatalo}, {Bacon}, {Blitz},
  {Bois}, {Bournaud}, {Bureau}, {Davies}, {Davis}, {de Zeeuw}, {Duc},
  {Khochfar}, {Kuntschner}, {Lablanche}, {Morganti}, {Naab}, {Oosterloo},
  {Sarzi}, {Serra}, \& {Weijmans}}]{cappellari11}
{Cappellari}, M., {Emsellem}, E., {Krajnovi{\'c}}, D., {et~al.} 2011, \mnras,
  413, 813

\bibitem[{{Cappellari} {et~al.}(2012){Cappellari}, {McDermid}, {Alatalo},
  {Blitz}, {Bois}, {Bournaud}, {Bureau}, {Crocker}, {Davies}, {Davis}, {de
  Zeeuw}, {Duc}, {Emsellem}, {Khochfar}, {Krajnovi{\'c}}, {Kuntschner},
  {Lablanche}, {Morganti}, {Naab}, {Oosterloo}, {Sarzi}, {Scott}, {Serra},
  {Weijmans}, \& {Young}}]{cappellari12}
{Cappellari}, M., {McDermid}, R.~M., {Alatalo}, K., {et~al.} 2012, \nat, 484,
  485

\bibitem[{{Cheung} {et~al.}(2016){Cheung}, {Bundy}, {Cappellari}, {Peirani},
  {Rujopakarn}, {Westfall}, {Yan}, {Bershady}, {Greene}, {Heckman}, {Drory},
  {Law}, {Masters}, {Thomas}, {Wake}, {Weijmans}, {Rubin}, {Belfiore},
  {Vulcani}, {Chen}, {Zhang}, {Gelfand}, {Bizyaev}, {Roman-Lopes}, \&
  {Schneider}}]{cheung16}
{Cheung}, E., {Bundy}, K., {Cappellari}, M., {et~al.} 2016, \nat, 533, 504

\bibitem[{{Ciotti} {et~al.}(2016){Ciotti}, {Pellegrini}, {Negri}, \&
  {Ostriker}}]{ciotti16}
{Ciotti}, L., {Pellegrini}, S., {Negri}, A., \& {Ostriker}, J.~P. 2016, ArXiv
  e-prints, arXiv:1608.03403

\bibitem[{{Combes} {et~al.}(2007){Combes}, {Young}, \& {Bureau}}]{combes07}
{Combes}, F., {Young}, L.~M., \& {Bureau}, M. 2007, \mnras, 377, 1795

\bibitem[{{Conroy} \& {van Dokkum}(2012)}]{conroy12}
{Conroy}, C., \& {van Dokkum}, P.~G. 2012, \apj, 760, 71

\bibitem[{{Crook} {et~al.}(2007){Crook}, {Huchra}, {Martimbeau}, {Masters},
  {Jarrett}, \& {Macri}}]{crook07}
{Crook}, A.~C., {Huchra}, J.~P., {Martimbeau}, N., {et~al.} 2007, \apj, 655,
  790

\bibitem[{{Crook} {et~al.}(2008){Crook}, {Huchra}, {Martimbeau}, {Masters},
  {Jarrett}, \& {Macri}}]{crook08}
---. 2008, \apj, 685, 1320

\bibitem[{{Davis} {et~al.}(2016){Davis}, {Greene}, {Ma}, {Pandya}, {Blakeslee},
  {McConnell}, \& {Thomas}}]{davis16a}
{Davis}, T.~A., {Greene}, J., {Ma}, C.-P., {et~al.} 2016, \mnras, 455, 214

\bibitem[{{Davis} {et~al.}(2011){Davis}, {Alatalo}, {Sarzi}, {Bureau}, {Young},
  {Blitz}, {Serra}, {Crocker}, {Krajnovi{\'c}}, {McDermid}, {Bois}, {Bournaud},
  {Cappellari}, {Davies}, {Duc}, {de Zeeuw}, {Emsellem}, {Khochfar},
  {Kuntschner}, {Lablanche}, {Morganti}, {Naab}, {Oosterloo}, {Scott}, \&
  {Weijmans}}]{davis11}
{Davis}, T.~A., {Alatalo}, K., {Sarzi}, M., {et~al.} 2011, \mnras, 417, 882

\bibitem[{{Davis} {et~al.}(2012){Davis}, {Krajnovi{\'c}}, {McDermid}, {Bureau},
  {Sarzi}, {Nyland}, {Alatalo}, {Bayet}, {Blitz}, {Bois}, {Bournaud},
  {Cappellari}, {Crocker}, {Davies}, {de Zeeuw}, {Duc}, {Emsellem}, {Khochfar},
  {Kuntschner}, {Lablanche}, {Morganti}, {Naab}, {Oosterloo}, {Scott}, {Serra},
  {Weijmans}, \& {Young}}]{davis12}
{Davis}, T.~A., {Krajnovi{\'c}}, D., {McDermid}, R.~M., {et~al.} 2012, \mnras,
  426, 1574

\bibitem[{{de Zeeuw} {et~al.}(2002){de Zeeuw}, {Bureau}, {Emsellem}, {Bacon},
  {Carollo}, {Copin}, {Davies}, {Kuntschner}, {Miller}, {Monnet}, {Peletier},
  \& {Verolme}}]{dezeeuw02}
{de Zeeuw}, P.~T., {Bureau}, M., {Emsellem}, E., {et~al.} 2002, \mnras, 329,
  513

\bibitem[{{Demoulin-Ulrich} {et~al.}(1984){Demoulin-Ulrich}, {Butcher}, \&
  {Boksenberg}}]{demoulinulrich84}
{Demoulin-Ulrich}, M.-H., {Butcher}, H.~R., \& {Boksenberg}, A. 1984, \apj,
  285, 527

\bibitem[{{Dopita} {et~al.}(2000){Dopita}, {Kewley}, {Heisler}, \&
  {Sutherland}}]{dopita00}
{Dopita}, M.~A., {Kewley}, L.~J., {Heisler}, C.~A., \& {Sutherland}, R.~S.
  2000, \apj, 542, 224

\bibitem[{{Dopita} \& {Sutherland}(1995)}]{ds95}
{Dopita}, M.~A., \& {Sutherland}, R.~S. 1995, \apj, 455, 468

\bibitem[{{Emsellem} {et~al.}(2007){Emsellem}, {Cappellari}, {Krajnovi{\'c}},
  {van de Ven}, {Bacon}, {Bureau}, {Davies}, {de Zeeuw}, {Falc{\'o}n-Barroso},
  {Kuntschner}, {McDermid}, {Peletier}, \& {Sarzi}}]{emsellem07}
{Emsellem}, E., {Cappellari}, M., {Krajnovi{\'c}}, D., {et~al.} 2007, \mnras,
  379, 401

\bibitem[{{Emsellem} {et~al.}(2011){Emsellem}, {Cappellari}, {Krajnovi{\'c}},
  {Alatalo}, {Blitz}, {Bois}, {Bournaud}, {Bureau}, {Davies}, {Davis}, {de
  Zeeuw}, {Khochfar}, {Kuntschner}, {Lablanche}, {McDermid}, {Morganti},
  {Naab}, {Oosterloo}, {Sarzi}, {Scott}, {Serra}, {van de Ven}, {Weijmans}, \&
  {Young}}]{emsellem11}
---. 2011, \mnras, 414, 888

\bibitem[{{Eracleous} {et~al.}(2010){Eracleous}, {Hwang}, \&
  {Flohic}}]{eracleous10}
{Eracleous}, M., {Hwang}, J.~A., \& {Flohic}, H.~M.~L.~G. 2010, \apj, 711, 796

\bibitem[{{Fabian}(1994)}]{fabian94}
{Fabian}, A.~C. 1994, \araa, 32, 277

\bibitem[{{Ferrarese} \& {Ford}(1999)}]{ferrarese99}
{Ferrarese}, L., \& {Ford}, H.~C. 1999, \apj, 515, 583

\bibitem[{{Forman} {et~al.}(1985){Forman}, {Jones}, \& {Tucker}}]{forman85}
{Forman}, W., {Jones}, C., \& {Tucker}, W. 1985, \apj, 293, 102

\bibitem[{{Gerhard}(1993)}]{gerhard93}
{Gerhard}, O.~E. 1993, \mnras, 265, 213

\bibitem[{{Gomes} {et~al.}(2016{\natexlab{a}}){Gomes}, {Papaderos},
  {V{\'{\i}}lchez}, {Kehrig}, {Iglesias-P{\'a}ramo}, {Breda}, {Lehnert},
  {S{\'a}nchez}, {Ziegler}, {Dos Reis}, {Bland-Hawthorn}, {Galbany}, {Bomans},
  {Rosales-Ortega}, {Walcher}, {Garc{\'{\i}}a-Benito}, {M{\'a}rquez}, {Del
  Olmo}, {Moll{\'a}}, {Marino}, {Catal{\'a}n-Torrecilla}, {Gonz{\'a}lez
  Delgado}, {L{\'o}pez-S{\'a}nchez}, \& {Califa Collaboration}}]{gomes16b}
{Gomes}, J.~M., {Papaderos}, P., {V{\'{\i}}lchez}, J.~M., {et~al.}
  2016{\natexlab{a}}, \aap, 586, A22

\bibitem[{{Gomes} {et~al.}(2016{\natexlab{b}}){Gomes}, {Papaderos}, {Kehrig},
  {V{\'{\i}}lchez}, {Lehnert}, {S{\'a}nchez}, {Ziegler}, {Breda}, {Dos Reis},
  {Iglesias-P{\'a}ramo}, {Bland-Hawthorn}, {Galbany}, {Bomans},
  {Rosales-Ortega}, {Cid Fernandes}, {Walcher}, {Falc{\'o}n-Barroso},
  {Garc{\'{\i}}a-Benito}, {M{\'a}rquez}, {Del Olmo}, {Masegosa}, {Moll{\'a}},
  {Marino}, {Gonz{\'a}lez Delgado}, {L{\'o}pez-S{\'a}nchez}, \& {Califa
  Collaboration}}]{gomes16}
{Gomes}, J.~M., {Papaderos}, P., {Kehrig}, C., {et~al.} 2016{\natexlab{b}},
  \aap, 588, A68

\bibitem[{{Goudfrooij}(1999)}]{goudfrooij99}
{Goudfrooij}, P. 1999, in Astronomical Society of the Pacific Conference
  Series, Vol. 163, Star Formation in Early Type Galaxies, ed. P.~{Carral} \&
  J.~{Cepa}, 55

\bibitem[{{Goudfrooij} {et~al.}(1994){Goudfrooij}, {Hansen}, {Jorgensen}, \&
  {Norgaard-Nielsen}}]{goudfrooij94}
{Goudfrooij}, P., {Hansen}, L., {Jorgensen}, H.~E., \& {Norgaard-Nielsen},
  H.~U. 1994, \aaps, 105

\bibitem[{{Goulding} {et~al.}(2016){Goulding}, {Greene}, {Ma}, {Veale},
  {Bogdan}, {Nyland}, {Blakeslee}, {McConnell}, \& {Thomas}}]{goulding16}
{Goulding}, A.~D., {Greene}, J.~E., {Ma}, C.-P., {et~al.} 2016, \apj, 826, 167

\bibitem[{{Graves} {et~al.}(2007){Graves}, {Faber}, {Schiavon}, \&
  {Yan}}]{graves07}
{Graves}, G.~J., {Faber}, S.~M., {Schiavon}, R.~P., \& {Yan}, R. 2007, \apj,
  671, 243

\bibitem[{{Greene} {et~al.}(2015){Greene}, {Janish}, {Ma}, {McConnell},
  {Blakeslee}, {Thomas}, \& {Murphy}}]{greene15}
{Greene}, J.~E., {Janish}, R., {Ma}, C.-P., {et~al.} 2015, \apj, 807, 11

\bibitem[{{Greene} {et~al.}(2012){Greene}, {Murphy}, {Comerford}, {Gebhardt},
  \& {Adams}}]{greene12}
{Greene}, J.~E., {Murphy}, J.~D., {Comerford}, J.~M., {Gebhardt}, K., \&
  {Adams}, J.~J. 2012, \apj, 750, 32

\bibitem[{{Groves} \& {Allen}(2010)}]{groves10}
{Groves}, B.~A., \& {Allen}, M.~G. 2010, \na, 15, 614

\bibitem[{{Groves} {et~al.}(2004{\natexlab{a}}){Groves}, {Dopita}, \&
  {Sutherland}}]{groves04a}
{Groves}, B.~A., {Dopita}, M.~A., \& {Sutherland}, R.~S. 2004{\natexlab{a}},
  \apjs, 153, 9

\bibitem[{{Groves} {et~al.}(2004{\natexlab{b}}){Groves}, {Dopita}, \&
  {Sutherland}}]{groves04b}
---. 2004{\natexlab{b}}, \apjs, 153, 75

\bibitem[{{Heckman}(1980)}]{heckman80}
{Heckman}, T.~M. 1980, \aap, 87, 152

\bibitem[{{Heckman} {et~al.}(1989){Heckman}, {Baum}, {van Breugel}, \&
  {McCarthy}}]{heckman89}
{Heckman}, T.~M., {Baum}, S.~A., {van Breugel}, W.~J.~M., \& {McCarthy}, P.
  1989, \apj, 338, 48

\bibitem[{{Ho}(2008)}]{ho08}
{Ho}, L.~C. 2008, \araa, 46, 475

\bibitem[{{Ho} {et~al.}(1997){Ho}, {Filippenko}, \& {Sargent}}]{ho97}
{Ho}, L.~C., {Filippenko}, A.~V., \& {Sargent}, W.~L.~W. 1997, \apjs, 112, 315

\bibitem[{{Huchra} {et~al.}(2012){Huchra}, {Macri}, {Masters}, {Jarrett},
  {Berlind}, {Calkins}, {Crook}, {Cutri}, {Erdo{\v g}du}, {Falco}, {George},
  {Hutcheson}, {Lahav}, {Mader}, {Mink}, {Martimbeau}, {Schneider},
  {Skrutskie}, {Tokarz}, \& {Westover}}]{huchra12}
{Huchra}, J.~P., {Macri}, L.~M., {Masters}, K.~L., {et~al.} 2012, \apjs, 199,
  26

\bibitem[{{Hudson} {et~al.}(2010){Hudson}, {Mittal}, {Reiprich}, {Nulsen},
  {Andernach}, \& {Sarazin}}]{hudson10}
{Hudson}, D.~S., {Mittal}, R., {Reiprich}, T.~H., {et~al.} 2010, \aap, 513, A37

\bibitem[{{Johansson} {et~al.}(2016){Johansson}, {Woods}, {Gilfanov}, {Sarzi},
  {Chen}, \& {Oh}}]{johansson16}
{Johansson}, J., {Woods}, T.~E., {Gilfanov}, M., {et~al.} 2016, \mnras, 461,
  4505

\bibitem[{{Kauffmann} {et~al.}(2003){Kauffmann}, {Heckman}, {White}, \&
  {Charlot}}]{kauffmann03}
{Kauffmann}, G., {Heckman}, T.~M., {White}, S.~D.~M., \& {Charlot}, S. 2003,
  \mnras, 341, 54

\bibitem[{{Kaviraj} {et~al.}(2007){Kaviraj}, {Schawinski}, {Devriendt},
  {Ferreras}, {Khochfar}, {Yoon}, {Yi}, {Deharveng}, {Boselli}, {Barlow},
  {Conrow}, {Forster}, {Friedman}, {Martin}, {Morrissey}, {Neff},
  {Schiminovich}, {Seibert}, {Small}, {Wyder}, {Bianchi}, {Donas}, {Heckman},
  {Lee}, {Madore}, {Milliard}, {Rich}, \& {Szalay}}]{kaviraj07}
{Kaviraj}, S., {Schawinski}, K., {Devriendt}, J.~E.~G., {et~al.} 2007, \apjs,
  173, 619

\bibitem[{{Kennicutt, R.}(1998)}]{kennicutt98}
{Kennicutt, R.} 1998, ARAA, 36

\bibitem[{{Kere{\v s}} {et~al.}(2005){Kere{\v s}}, {Katz}, {Weinberg}, \&
  {Dav{\'e}}}]{keres05}
{Kere{\v s}}, D., {Katz}, N., {Weinberg}, D.~H., \& {Dav{\'e}}, R. 2005,
  \mnras, 363, 2

\bibitem[{{Kewley} \& {Dopita}(2002)}]{kewley02}
{Kewley}, L.~J., \& {Dopita}, M.~A. 2002, \apjs, 142, 35

\bibitem[{{Kewley} {et~al.}(2006){Kewley}, {Groves}, {Kauffmann}, \&
  {Heckman}}]{kewley06}
{Kewley}, L.~J., {Groves}, B., {Kauffmann}, G., \& {Heckman}, T. 2006, \mnras,
  372, 961

\bibitem[{{Khosroshahi} {et~al.}(2004){Khosroshahi}, {Jones}, \&
  {Ponman}}]{khosroshahi04}
{Khosroshahi}, H.~G., {Jones}, L.~R., \& {Ponman}, T.~J. 2004, \mnras, 349,
  1240

\bibitem[{{Kim}(1989)}]{kim89}
{Kim}, D.-W. 1989, \apj, 346, 653

\bibitem[{{Knapp} \& {Rupen}(1996)}]{knapp96b}
{Knapp}, G.~R., \& {Rupen}, M.~P. 1996, \apj, 460, 271

\bibitem[{{Knapp} {et~al.}(1996){Knapp}, {Rupen}, {Fich}, {Harper}, \&
  {Wynn-Williams}}]{knapp96}
{Knapp}, G.~R., {Rupen}, M.~P., {Fich}, M., {Harper}, D.~A., \&
  {Wynn-Williams}, C.~G. 1996, \aap, 315, L75

\bibitem[{{Knapp} {et~al.}(1985){Knapp}, {Turner}, \& {Cunniffe}}]{knapp85}
{Knapp}, G.~R., {Turner}, E.~L., \& {Cunniffe}, P.~E. 1985, \aj, 90, 454

\bibitem[{{Koleva} {et~al.}(2009){Koleva}, {Prugniel}, {Bouchard}, \&
  {Wu}}]{koleva09}
{Koleva}, M., {Prugniel}, P., {Bouchard}, A., \& {Wu}, Y. 2009, \aap, 501, 1269

\bibitem[{{Kormendy} \& {Bender}(2009)}]{kormendy09}
{Kormendy}, J., \& {Bender}, R. 2009, \apjl, 691, L142

\bibitem[{{Koski} \& {Osterbrock}(1976)}]{koski76}
{Koski}, A.~T., \& {Osterbrock}, D.~E. 1976, \apjl, 203, L49

\bibitem[{{Krajnovi{\'c}} {et~al.}(2006){Krajnovi{\'c}}, {Cappellari}, {de
  Zeeuw}, \& {Copin}}]{krajnovic06}
{Krajnovi{\'c}}, D., {Cappellari}, M., {de Zeeuw}, P.~T., \& {Copin}, Y. 2006,
  \mnras, 366, 787

\bibitem[{{Krajnovi{\'c}} {et~al.}(2011){Krajnovi{\'c}}, {Emsellem},
  {Cappellari}, {Alatalo}, {Blitz}, {Bois}, {Bournaud}, {Bureau}, {Davies},
  {Davis}, {de Zeeuw}, {Khochfar}, {Kuntschner}, {Lablanche}, {McDermid},
  {Morganti}, {Naab}, {Oosterloo}, {Sarzi}, {Scott}, {Serra}, {Weijmans}, \&
  {Young}}]{krajnovic11}
{Krajnovi{\'c}}, D., {Emsellem}, E., {Cappellari}, M., {et~al.} 2011, \mnras,
  414, 2923

\bibitem[{{Lagos} {et~al.}(2015){Lagos}, {Padilla}, {Davis}, {Lacey}, {Baugh},
  {Gonzalez-Perez}, {Zwaan}, \& {Contreras}}]{lagos15}
{Lagos}, C.~d.~P., {Padilla}, N.~D., {Davis}, T.~A., {et~al.} 2015, \mnras,
  448, 1271

\bibitem[{{Leitherer} {et~al.}(1999){Leitherer}, {Schaerer}, {Goldader},
  {Delgado}, {Robert}, {Kune}, {de Mello}, {Devost}, \&
  {Heckman}}]{leitherer99}
{Leitherer}, C., {Schaerer}, D., {Goldader}, J.~D., {et~al.} 1999, \apjs, 123,
  3

\bibitem[{{Ma} {et~al.}(2014){Ma}, {Greene}, {McConnell}, {Janish},
  {Blakeslee}, {Thomas}, \& {Murphy}}]{ma14}
{Ma}, C.-P., {Greene}, J.~E., {McConnell}, N., {et~al.} 2014, \apj, 795, 158

\bibitem[{{Macchetto} {et~al.}(1996){Macchetto}, {Pastoriza}, {Caon}, {Sparks},
  {Giavalisco}, {Bender}, \& {Capaccioli}}]{macchetto96}
{Macchetto}, F., {Pastoriza}, M., {Caon}, N., {et~al.} 1996, \aaps, 120, 463

\bibitem[{{Martel} {et~al.}(2004){Martel}, {Ford}, {Bradley}, {Tran},
  {Menanteau}, {Tsvetanov}, {Illingworth}, {Hartig}, \& {Clampin}}]{martel04}
{Martel}, A.~R., {Ford}, H.~C., {Bradley}, L.~D., {et~al.} 2004, \aj, 128, 2758

\bibitem[{{Martin} {et~al.}(2005){Martin}, {Fanson}, {Schiminovich},
  {Morrissey}, {Friedman}, {Barlow}, {Conrow}, {Grange}, {Jelinsky},
  {Milliard}, {Siegmund}, {Bianchi}, {Byun}, {Donas}, {Forster}, {Heckman},
  {Lee}, {Madore}, {Malina}, {Neff}, {Rich}, {Small}, {Surber}, {Szalay},
  {Welsh}, \& {Wyder}}]{martin05}
{Martin}, D.~C., {Fanson}, J., {Schiminovich}, D., {et~al.} 2005, \apjl, 619,
  L1

\bibitem[{{Martini} {et~al.}(2013){Martini}, {Dicken}, \&
  {Storchi-Bergmann}}]{martini13}
{Martini}, P., {Dicken}, D., \& {Storchi-Bergmann}, T. 2013, \apj, 766, 121

\bibitem[{{Martini} {et~al.}(2003){Martini}, {Regan}, {Mulchaey}, \&
  {Pogge}}]{martini03b}
{Martini}, P., {Regan}, M.~W., {Mulchaey}, J.~S., \& {Pogge}, R.~W. 2003, \apj,
  589, 774

\bibitem[{{Mathews} \& {Brighenti}(2003)}]{mathews03}
{Mathews}, W.~G., \& {Brighenti}, F. 2003, \araa, 41, 191

\bibitem[{{McDonald} {et~al.}(2011){McDonald}, {Veilleux}, \&
  {Mushotzky}}]{mcdonald11a}
{McDonald}, M., {Veilleux}, S., \& {Mushotzky}, R. 2011, \apj, 731, 33

\bibitem[{{McDonald} {et~al.}(2010){McDonald}, {Veilleux}, {Rupke}, \&
  {Mushotzky}}]{mcdonald10}
{McDonald}, M., {Veilleux}, S., {Rupke}, D.~S.~N., \& {Mushotzky}, R. 2010,
  \apj, 721, 1262

\bibitem[{{Moustakas} \& {Kennicutt}(2006)}]{moustakas06}
{Moustakas}, J., \& {Kennicutt}, Jr., R.~C. 2006, \apj, 651, 155

\bibitem[{{Mulchaey}(2000)}]{mulchaey00}
{Mulchaey}, J.~S. 2000, \araa, 38, 289

\bibitem[{{Naab} {et~al.}(2014){Naab}, {Oser}, {Emsellem}, {Cappellari},
  {Krajnovi{\'c}}, {McDermid}, {Alatalo}, {Bayet}, {Blitz}, {Bois}, {Bournaud},
  {Bureau}, {Crocker}, {Davies}, {Davis}, {de Zeeuw}, {Duc}, {Hirschmann},
  {Johansson}, {Khochfar}, {Kuntschner}, {Morganti}, {Oosterloo}, {Sarzi},
  {Scott}, {Serra}, {Ven}, {Weijmans}, \& {Young}}]{naab14}
{Naab}, T., {Oser}, L., {Emsellem}, E., {et~al.} 2014, \mnras, 444, 3357

\bibitem[{{Nesvadba} {et~al.}(2011){Nesvadba}, {Polletta}, {Lehnert},
  {Bergeron}, {De Breuck}, {Lagache}, \& {Omont}}]{nesvadba11}
{Nesvadba}, N.~P.~H., {Polletta}, M., {Lehnert}, M.~D., {et~al.} 2011, \mnras,
  415, 2359

\bibitem[{{Nyland} {et~al.}(2013){Nyland}, {Alatalo}, {Wrobel}, {Young},
  {Morganti}, {Davis}, {de Zeeuw}, {Deustua}, \& {Bureau}}]{nyland13}
{Nyland}, K., {Alatalo}, K., {Wrobel}, J.~M., {et~al.} 2013, \apj, 779, 173

\bibitem[{{Nyland} {et~al.}(2016){Nyland}, {Young}, {Wrobel}, {Sarzi},
  {Morganti}, {Alatalo}, {Blitz}, {Bournaud}, {Bureau}, {Cappellari},
  {Crocker}, {Davies}, {Davis}, {de Zeeuw}, {Duc}, {Emsellem}, {Khochfar},
  {Krajnovi{\'c}}, {Kuntschner}, {McDermid}, {Naab}, {Oosterloo}, {Scott},
  {Serra}, \& {Weijmans}}]{nyland16}
{Nyland}, K., {Young}, L.~M., {Wrobel}, J.~M., {et~al.} 2016, \mnras, 458, 2221

\bibitem[{{Osterbrock}(1989)}]{osterbrock89}
{Osterbrock}, D.~E. 1989, {Astrophysics of gaseous nebulae and active galactic
  nuclei}

\bibitem[{{O'Sullivan} {et~al.}(2015){O'Sullivan}, {Combes}, {Hamer},
  {Salom{\'e}}, {Babul}, \& {Raychaudhury}}]{osullivan15}
{O'Sullivan}, E., {Combes}, F., {Hamer}, S., {et~al.} 2015, \aap, 573, A111

\bibitem[{{Papaderos} {et~al.}(2013){Papaderos}, {Gomes}, {V{\'{\i}}lchez},
  {Kehrig}, {Lehnert}, {Ziegler}, {S{\'a}nchez}, {Husemann}, {Monreal-Ibero},
  {Garc{\'{\i}}a-Benito}, {Bland-Hawthorn}, {Cortijo-Ferrero}, {de
  Lorenzo-C{\'a}ceres}, {del Olmo}, {Falc{\'o}n-Barroso}, {Galbany},
  {Iglesias-P{\'a}ramo}, {L{\'o}pez-S{\'a}nchez}, {Marquez}, {Moll{\'a}},
  {Mast}, {van de Ven}, \& {Wisotzki}}]{papaderos13}
{Papaderos}, P., {Gomes}, J.~M., {V{\'{\i}}lchez}, J.~M., {et~al.} 2013, \aap,
  555, L1

\bibitem[{{Paturel} {et~al.}(2003){Paturel}, {Petit}, {Prugniel}, {Theureau},
  {Rousseau}, {Brouty}, {Dubois}, \& {Cambr{\'e}sy}}]{paturel03}
{Paturel}, G., {Petit}, C., {Prugniel}, P., {et~al.} 2003, \aap, 412, 45

\bibitem[{{Phillips} {et~al.}(1986){Phillips}, {Jenkins}, {Dopita}, {Sadler},
  \& {Binette}}]{phillips86}
{Phillips}, M.~M., {Jenkins}, C.~R., {Dopita}, M.~A., {Sadler}, E.~M., \&
  {Binette}, L. 1986, \aj, 91, 1062

\bibitem[{{Salviander} {et~al.}(2007){Salviander}, {Shields}, {Gebhardt}, \&
  {Bonning}}]{salviander07}
{Salviander}, S., {Shields}, G.~A., {Gebhardt}, K., \& {Bonning}, E.~W. 2007,
  \apj, 662, 131

\bibitem[{{Sarzi} {et~al.}(2006){Sarzi}, {Falc{\'o}n-Barroso}, {Davies},
  {Bacon}, {Bureau}, {Cappellari}, {de Zeeuw}, {Emsellem}, {Fathi},
  {Krajnovi{\'c}}, {Kuntschner}, {McDermid}, \& {Peletier}}]{sarzi06}
{Sarzi}, M., {Falc{\'o}n-Barroso}, J., {Davies}, R.~L., {et~al.} 2006, \mnras,
  366, 1151

\bibitem[{{Sarzi} {et~al.}(2010){Sarzi}, {Shields}, {Schawinski}, {Jeong},
  {Shapiro}, {Bacon}, {Bureau}, {Cappellari}, {Davies}, {de Zeeuw}, {Emsellem},
  {Falc{\'o}n-Barroso}, {Krajnovi{\'c}}, {Kuntschner}, {McDermid}, {Peletier},
  {van den Bosch}, {van de Ven}, \& {Yi}}]{sarzi10}
{Sarzi}, M., {Shields}, J.~C., {Schawinski}, K., {et~al.} 2010, \mnras, 402,
  2187

\bibitem[{{Sarzi} {et~al.}(2013){Sarzi}, {Alatalo}, {Blitz}, {Bois},
  {Bournaud}, {Bureau}, {Cappellari}, {Crocker}, {Davies}, {Davis}, {de Zeeuw},
  {Duc}, {Emsellem}, {Khochfar}, {Krajnovi{\'c}}, {Kuntschner}, {Lablanche},
  {McDermid}, {Morganti}, {Naab}, {Oosterloo}, {Scott}, {Serra}, {Young}, \&
  {Weijmans}}]{sarzi13}
{Sarzi}, M., {Alatalo}, K., {Blitz}, L., {et~al.} 2013, \mnras, 432, 1845

\bibitem[{{Serra} {et~al.}(2014){Serra}, {Oser}, {Krajnovi{\'c}}, {Naab},
  {Oosterloo}, {Morganti}, {Cappellari}, {Emsellem}, {Young}, {Blitz}, {Davis},
  {Duc}, {Hirschmann}, {Weijmans}, {Alatalo}, {Bayet}, {Bois}, {Bournaud},
  {Bureau}, {Crocker}, {Davies}, {de Zeeuw}, {Khochfar}, {Kuntschner},
  {Lablanche}, {McDermid}, {Sarzi}, \& {Scott}}]{serra14}
{Serra}, P., {Oser}, L., {Krajnovi{\'c}}, D., {et~al.} 2014, \mnras, 444, 3388

\bibitem[{{Simonian} \& {Martini}(2016)}]{simonian16}
{Simonian}, G.~V., \& {Martini}, P. 2016, ArXiv e-prints, arXiv:1603.09345

\bibitem[{{Sparks} {et~al.}(1989){Sparks}, {Macchetto}, \&
  {Golombek}}]{sparks89}
{Sparks}, W.~B., {Macchetto}, F., \& {Golombek}, D. 1989, \apj, 345, 153

\bibitem[{{Stasi{\'n}ska} {et~al.}(2008){Stasi{\'n}ska}, {Vale Asari}, {Cid
  Fernandes}, {Gomes}, {Schlickmann}, {Mateus}, {Schoenell}, {Sodr{\'e}}, \&
  {Seagal Collaboration}}]{stasinska08}
{Stasi{\'n}ska}, G., {Vale Asari}, N., {Cid Fernandes}, R., {et~al.} 2008,
  \mnras, 391, L29

\bibitem[{{Struve} {et~al.}(2010){Struve}, {Oosterloo}, {Sancisi}, {Morganti},
  \& {Emonts}}]{struve10}
{Struve}, C., {Oosterloo}, T., {Sancisi}, R., {Morganti}, R., \& {Emonts},
  B.~H.~C. 2010, \aap, 523, A75

\bibitem[{{Temi} {et~al.}(2007){Temi}, {Brighenti}, \& {Mathews}}]{temi07}
{Temi}, P., {Brighenti}, F., \& {Mathews}, W.~G. 2007, \apj, 660, 1215

\bibitem[{{Thomas} {et~al.}(2016){Thomas}, {Ma}, {McConnell}, {Greene},
  {Blakeslee}, \& {Janish}}]{thomas16}
{Thomas}, J., {Ma}, C.-P., {McConnell}, N.~J., {et~al.} 2016, \nat, 532, 340

\bibitem[{{Tremonti} {et~al.}(2004){Tremonti}, {Heckman}, {Kauffmann},
  {Brinchmann}, {Charlot}, {White}, {Seibert}, {Peng}, {Schlegel}, {Uomoto},
  {Fukugita}, \& {Brinkmann}}]{tremonti04}
{Tremonti}, C.~A., {Heckman}, T.~M., {Kauffmann}, G., {et~al.} 2004, \apj, 613,
  898

\bibitem[{{Treu} {et~al.}(2010){Treu}, {Auger}, {Koopmans}, {Gavazzi},
  {Marshall}, \& {Bolton}}]{treu10}
{Treu}, T., {Auger}, M.~W., {Koopmans}, L.~V.~E., {et~al.} 2010, \apj, 709,
  1195

\bibitem[{{van de Voort} {et~al.}(2015){van de Voort}, {Davis}, {Kere{\v s}},
  {Quataert}, {Faucher-Gigu{\`e}re}, \& {Hopkins}}]{vandevoort15}
{van de Voort}, F., {Davis}, T.~A., {Kere{\v s}}, D., {et~al.} 2015, \mnras,
  451, 3269

\bibitem[{{van der Marel} \& {Franx}(1993)}]{vdmarel93}
{van der Marel}, R.~P., \& {Franx}, M. 1993, \apj, 407, 525

\bibitem[{{van Dokkum} \& {Franx}(1995)}]{vandokkum95}
{van Dokkum}, P.~G., \& {Franx}, M. 1995, \aj, 110, 2027

\bibitem[{{Veale} {et~al.}(2016){Veale}, {Ma}, {Thomas}, {Greene}, {McConnell},
  {Walsh}, {Ito}, {Blakeslee}, \& {Janish}}]{veale16}
{Veale}, M., {Ma}, C.-P., {Thomas}, J., {et~al.} 2016, ArXiv e-prints,
  arXiv:1609.00391

\bibitem[{{Veilleux} \& {Osterbrock}(1987)}]{veilleuxosterbrock87}
{Veilleux}, S., \& {Osterbrock}, D.~E. 1987, \apjs, 63, 295

\bibitem[{{Werner} {et~al.}(2014){Werner}, {Oonk}, {Sun}, {Nulsen}, {Allen},
  {Canning}, {Simionescu}, {Hoffer}, {Connor}, {Donahue}, {Edge}, {Fabian},
  {von der Linden}, {Reynolds}, \& {Ruszkowski}}]{werner14}
{Werner}, N., {Oonk}, J.~B.~R., {Sun}, M., {et~al.} 2014, \mnras, 439, 2291

\bibitem[{{Yan} \& {Blanton}(2012)}]{yan12}
{Yan}, R., \& {Blanton}, M.~R. 2012, \apj, 747, 61

\bibitem[{{Young} {et~al.}(2011){Young}, {Bureau}, {Davis}, {Combes},
  {McDermid}, {Alatalo}, {Blitz}, {Bois}, {Bournaud}, {Cappellari}, {Davies},
  {de Zeeuw}, {Emsellem}, {Khochfar}, {Krajnovi{\'c}}, {Kuntschner},
  {Lablanche}, {Morganti}, {Naab}, {Oosterloo}, {Sarzi}, {Scott}, {Serra}, \&
  {Weijmans}}]{young11}
{Young}, L.~M., {Bureau}, M., {Davis}, T.~A., {et~al.} 2011, \mnras, 414, 940

\bibitem[{{Zeilinger} {et~al.}(1996){Zeilinger}, {Pizzella}, {Amico}, {Bertin},
  {Bertola}, {Buson}, {Danziger}, {Dejonghe}, {Sadler}, {Saglia}, \& {de
  Zeeuw}}]{zeilinger96}
{Zeilinger}, W.~W., {Pizzella}, A., {Amico}, P., {et~al.} 1996, \aaps, 120, 257

\end{thebibliography}

\appendix

\section{Integrated Flux Completeness Limits from Simulations}\label{sec:completeness}
In this appendix, we present simulations that help to quantify the integrated flux completeness limits of our emission line detection algorithm. Our IFU spectra span a wide range of full continuum S/N ratios that vary with galactocentric radius and differ from galaxy to galaxy. The individual fibers closest to the photometric center have full continuum S/N $\gtrsim20$ (reaching above 100 for some galaxies), and there is a gradual drop with radius to values as low as $\sim1$ for the outermost individual fibers (but the binned spectra at these large galactocentric radii always have full continuum S/N $\gtrsim20$). Therefore, we created a large set of synthetic spectra assuming three different continuum S/N ratio regimes: low ($S/N=10$), intermediate ($S/N=40$), and high ($S/N=70$). The synthetic spectra were constructed as follows. We began with the central fiber spectrum of a random galaxy in our sample (NGC 2274) that does not have emission lines, and degraded its continuum $S/N$ to one of the aforementioned fixed values by scaling its error spectrum and adding Gaussian random noise per pixel. By using the central fiber spectrum of a real galaxy rather than a best-fit composite pPXF stellar template (which naturally does not have emission lines), we are to some degree able to include systematic observational errors in our simulations (as well as features in wavelength regions that current stellar population models are not able to reproduce; e.g., the [\ion{N}{1}] 5200\AA\ region). 

We then superimposed Gaussian emission lines assuming a range of input significance values: $\mathcal{S}=0.5,1.0,1.5,2.0,2.5,3.0,3.5,4.0,5.0,10.0$. With an assumed line significance level and a measure of the local continuum RMS level, we can solve for the input amplitude of a line ($A=N\times\mathcal{S}$). For a given combination of degraded continuum $S/N$ and input line significance $\mathcal{S}$, we created 100 realizations. Since we will use our simulations to only probe the detectability of integrated flux of emission lines, the velocity and velocity dispersion of simulated emission lines are all fixed to 0 and 200 km s$^{-1}$, respectively. Our simulations will test the detectability of the [\ion{O}{2}], [\ion{O}{3}] 4959,5007, and H$\beta$ emission lines -- we do not simulate the detectability of the [\ion{N}{1}] line because we adopt quite stringent significance criteria for it anyway (see \autoref{sec:emlinedet}). We note that our simulations are not suitable for studying systematic uncertainties due to stellar continuum modeling and subtraction on a galaxy-by-galaxy basis; rather, our simulations quantify the detectability of emission lines based on a sufficiently large and representative grid of local continuum S/N and line significance values.

\begin{figure*} 
\begin{center}
\includegraphics[width=\hsize]{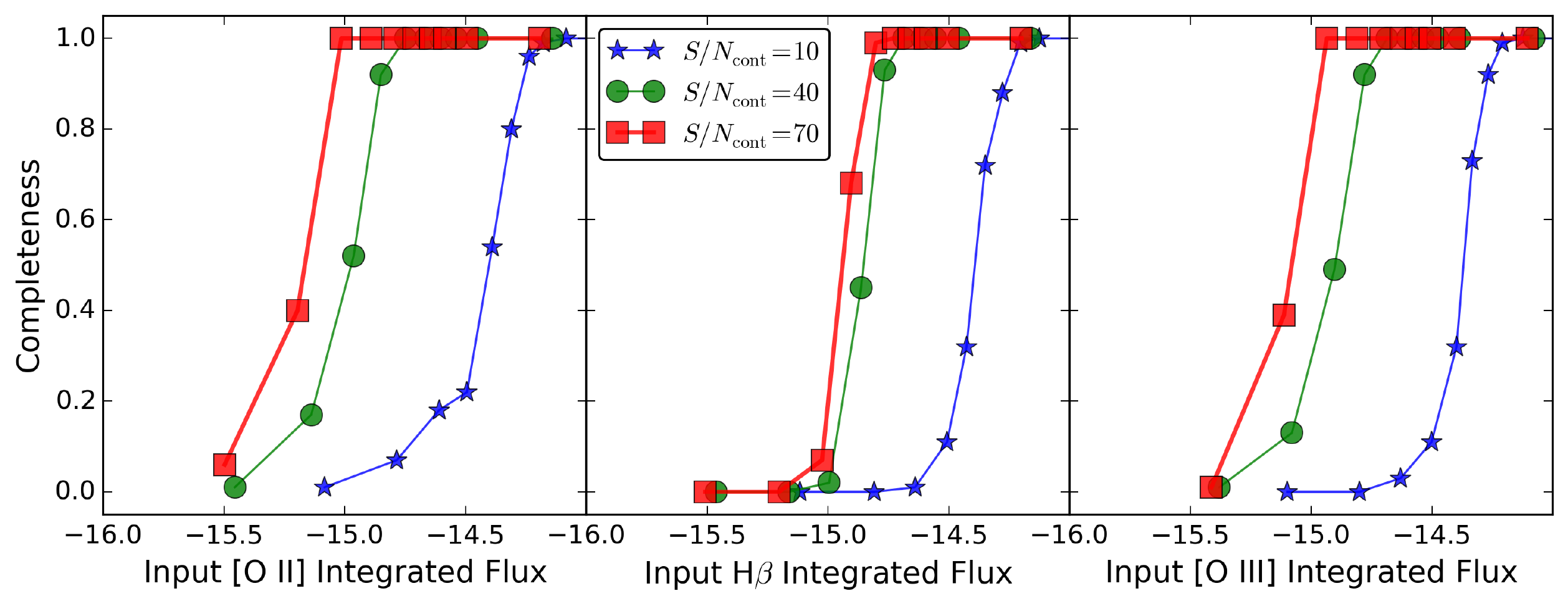}
\end{center}
\caption{Our completeness as a function of input integrated flux for the three emission lines considered in this paper: [\ion{O}{2}] (left), H$\beta$ (middle), and [\ion{O}{3}] 5007 (right). We present the completeness curves assuming three different continuum S/N ratios: 10 (blue stars/lines), 40 (green circles/lines), and 70 (red squares/lines). As expected, emission lines in noisier spectra need to be stronger to be significantly detected.}
\label{fig:completeness}
\end{figure*}

With the synthetic spectra in hand, we re-did our entire analysis on the simulated dataset in the same way as for the observations (running pPXF, subtracting the best-fit composite stellar template, fitting the residual emission lines, etc). \autoref{fig:completeness} shows our ability to recover the [\ion{O}{2}], H$\beta$ and [\ion{O}{3}] 5007 emission lines as a function of input integrated flux and degraded continuum S/N ratio. We find the sensible result that as the continuum S/N decreases, an emission line must become stronger for it to be significantly detected. In the highest continuum S/N regime, we can get down to about $3\times10^{-16}$ erg s$^{-1}$ cm$^{-2}$ integrated flux for [\ion{O}{2}], though fainter lines in the observed spectra are detected and can be visually confirmed. The integrated flux detectability thresholds for [\ion{O}{3}] are a bit higher than those of [\ion{O}{2}] probably for two reasons: (1) [\ion{O}{3}] 5007 is in general fainter than [\ion{O}{2}], and (2) in our red galaxies, the stellar continuum is much stronger in the [\ion{O}{3}] rather than [\ion{O}{2}] wavelength region. H$\beta$ has the highest completeness threshold because it is significantly fainter, is dependent on proper continuum subtraction, and is considered detected only if [\ion{O}{2}] is also already detected. 

The fact that our simulated [\ion{O}{2}] integrated fluxes get down to about $3\times10^{-16}$ erg s$^{-1}$ cm$^{-2}$ agrees well with the typical measurements from the observations. Considering NGC 1453 as a prototypical example, the integrated flux of [\ion{O}{2}] in the central fiber is $\sim7\times10^{-15}$ erg s$^{-1}$ cm$^{-2}$. In contrast, the integrated flux of [\ion{O}{2}] in the fiber near $R_{\rm gas}$ is only $\sim1.1\times10^{-17}$ erg s$^{-1}$ cm$^{-2}$. Upper limits on the integrated flux for [\ion{O}{2}] non-detections are all less than $3\times10^{-16}$ erg s$^{-1}$ cm$^{-2}$, with a median value of $\sim1.7\times10^{-16}$ erg s$^{-1}$ cm$^{-2}$. The integrated fluxes are recovered to within $5\%$ error (in the median) relative to the input value, and the line widths are recovered to within $3\%$ (in the median). We therefore assign all integrated flux measurements in the real data a conservative relative uncertainty of $10\%$. 

\section{Two-dimensional Maps of MASSIVE Galaxies with Warm Ionized Gas}\label{sec:maps2d}
\begin{figure*} 
\begin{center}
\includegraphics[width=\hsize]{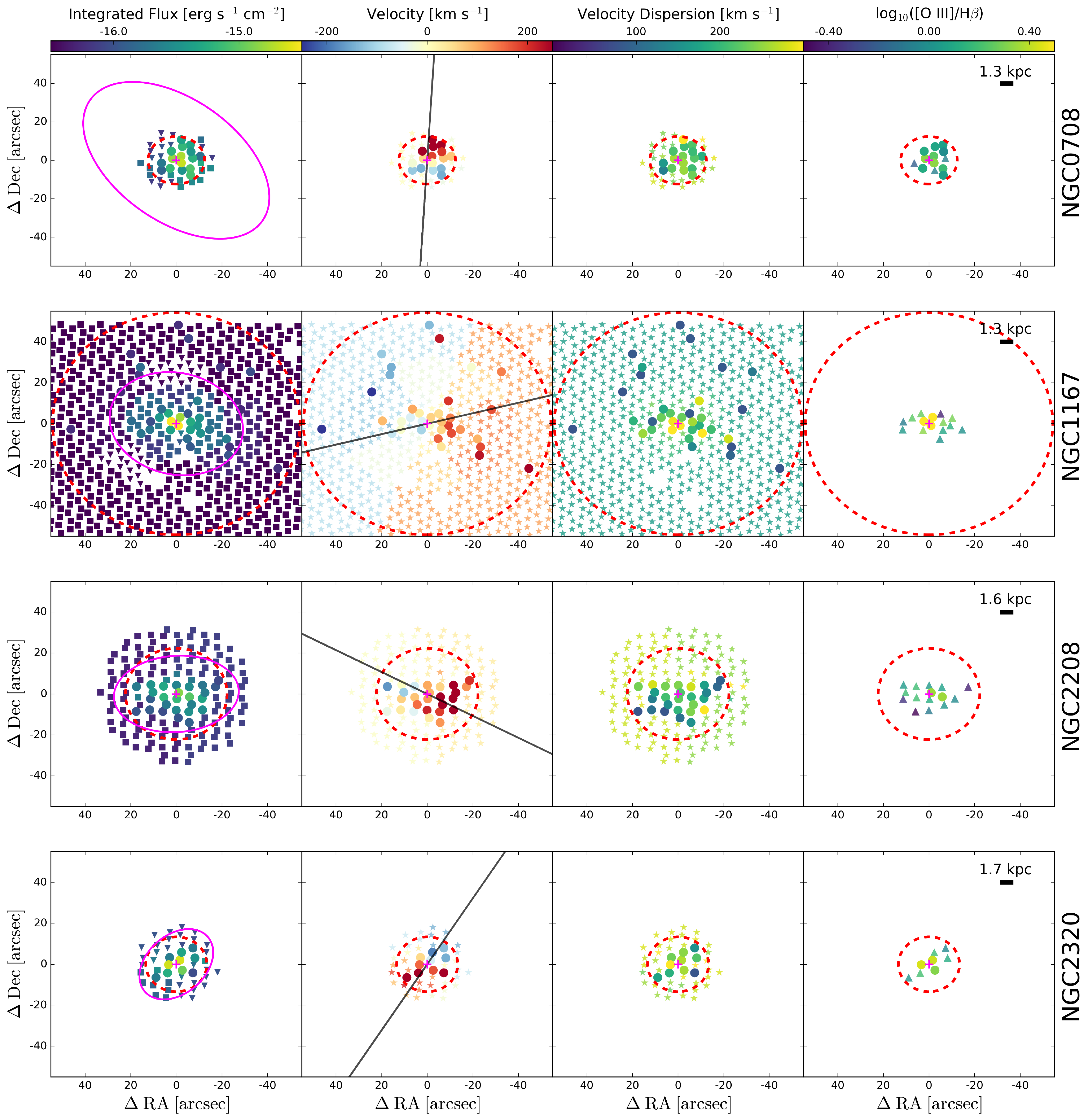}
\end{center}
\caption{Two-dimensional maps of various quantities for MASSIVE galaxies with detected warm ionized gas. From left to right: [\ion{O}{2}] integrated flux, gas and stellar velocity, gas and stellar velocity dispersion, and [\ion{O}{3}]/H$\beta$ integrated flux ratio. The pink cross in each subpanel marks the photometric center of the galaxy. The pink ellipse in the integrated flux subplot shows the stellar effective radius, photometric position angle, and axis ratio, whereas the red dashed circle in each panel marks the maximum radial extent of the gas, $R_{\rm gas}$. The black line in the velocity subplot shows the kinematic position angle of the rotating gas. In the velocity and velocity dispersion subplots, the star marker symbols in the background represent the velocities and velocity dispersions of the stars in each fiber as derived with pPXF. In all panels, circle marker symbols represent measurements from the individual fibers and squares represent measurements from the bins (shown where the single-fiber measurements are insignificant or have large uncertainties). In the integrated flux subplot, downward facing triangles represent upper limits, whereas in the excitation ratio subplot, upward facing triangles represent lower limits (shown where [\ion{O}{3}] is detected but H$\beta$ is not, leading to an H$\beta$ upper limit in the denominator). The number of kpc corresponding to four arcsec at the redshift of each galaxy is written in the rightmost subplot.}
\label{fig:portraits}
\end{figure*}

In this appendix, we show two-dimensional maps of various quantities for MASSIVE galaxies in which we detect warm ionized gas. The ordering of the galaxies in \autoref{fig:portraits} is the same as in \autoref{tab:det}: extended and rotating gas, extended but patchy gas, and unresolved compact gas (NGC 1453 is shown separately in \autoref{fig:portrait1453}). 

\begin{figure*} 
\begin{center}
\includegraphics[width=\hsize]{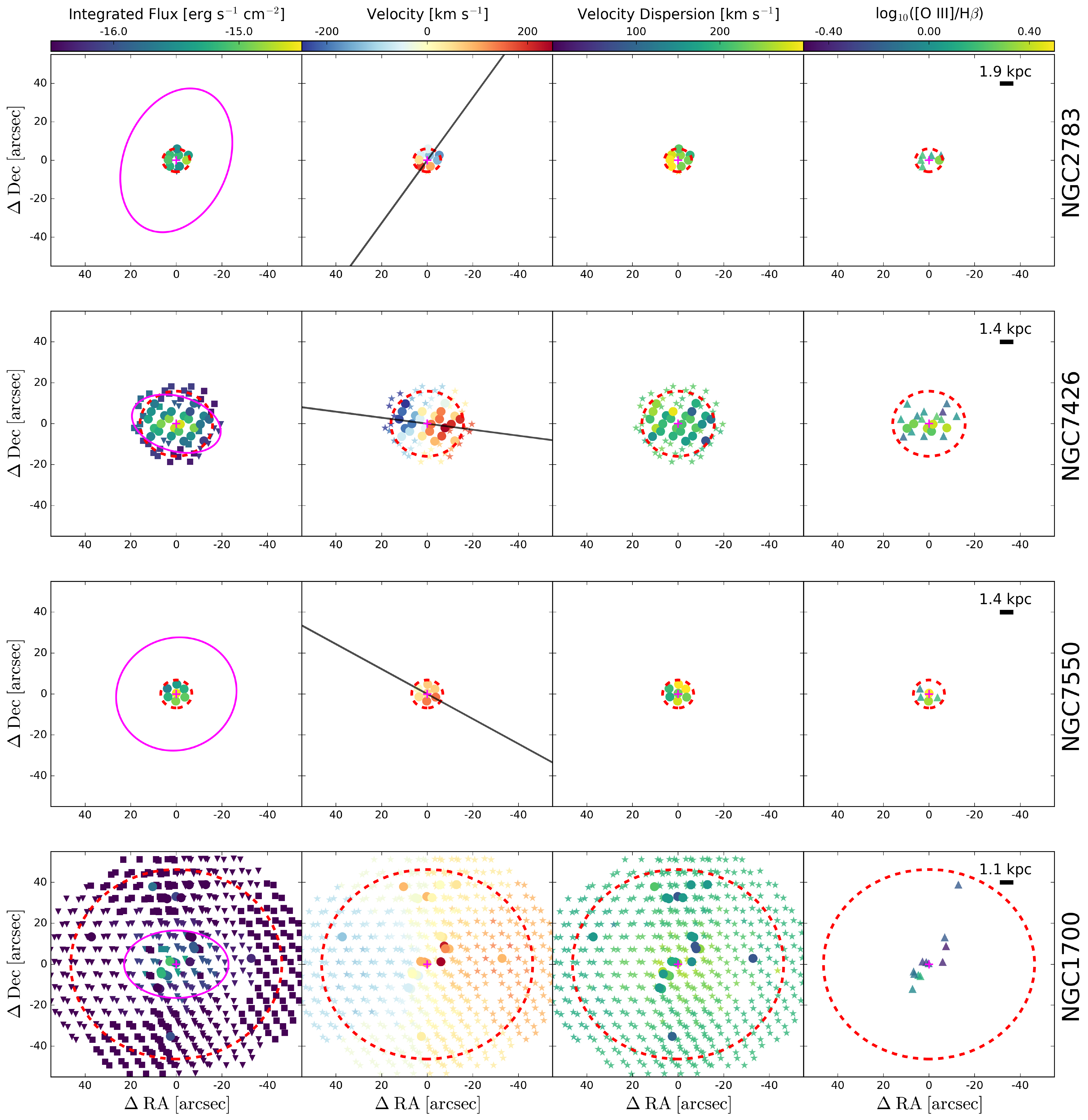}
\end{center}
\caption{Two-dimensional maps of various quantities for MASSIVE galaxies with detected warm ionized gas. From left to right: [\ion{O}{2}] integrated flux, gas and stellar velocity, gas and stellar velocity dispersion, and [\ion{O}{3}]/H$\beta$ integrated flux ratio. The plotting conventions, symbols and colors are the same as described in \autoref{fig:portraits}.}
\end{figure*}

\begin{figure*} 
\begin{center}
\includegraphics[width=\hsize]{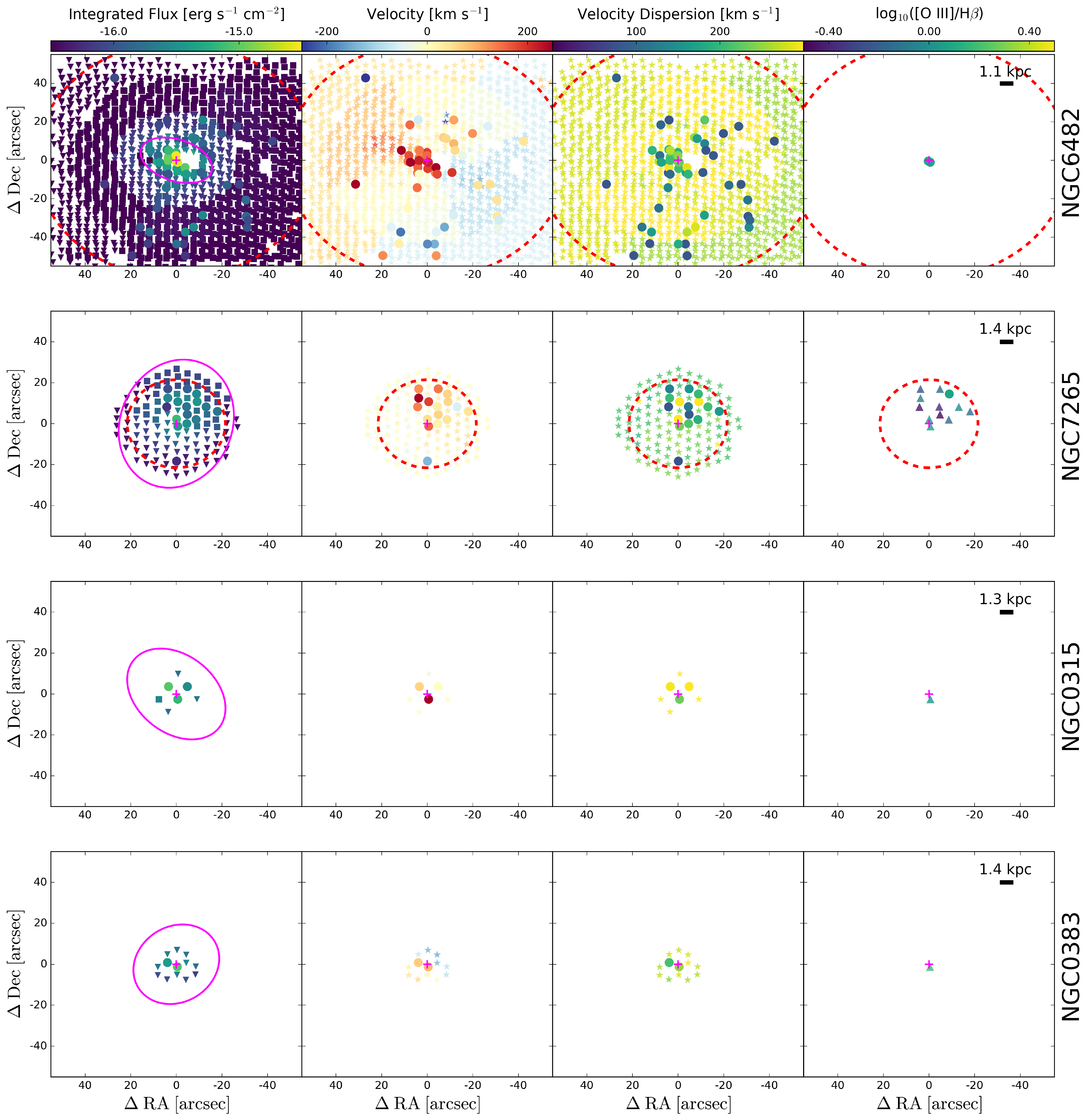}
\end{center}
\caption{Two-dimensional maps of various quantities for MASSIVE galaxies with detected warm ionized gas. From left to right: [\ion{O}{2}] integrated flux, gas and stellar velocity, gas and stellar velocity dispersion, and [\ion{O}{3}]/H$\beta$ integrated flux ratio. The plotting conventions, symbols and colors are the same as described in \autoref{fig:portraits}.}
\end{figure*}

\begin{figure*} 
\begin{center}
\includegraphics[width=\hsize]{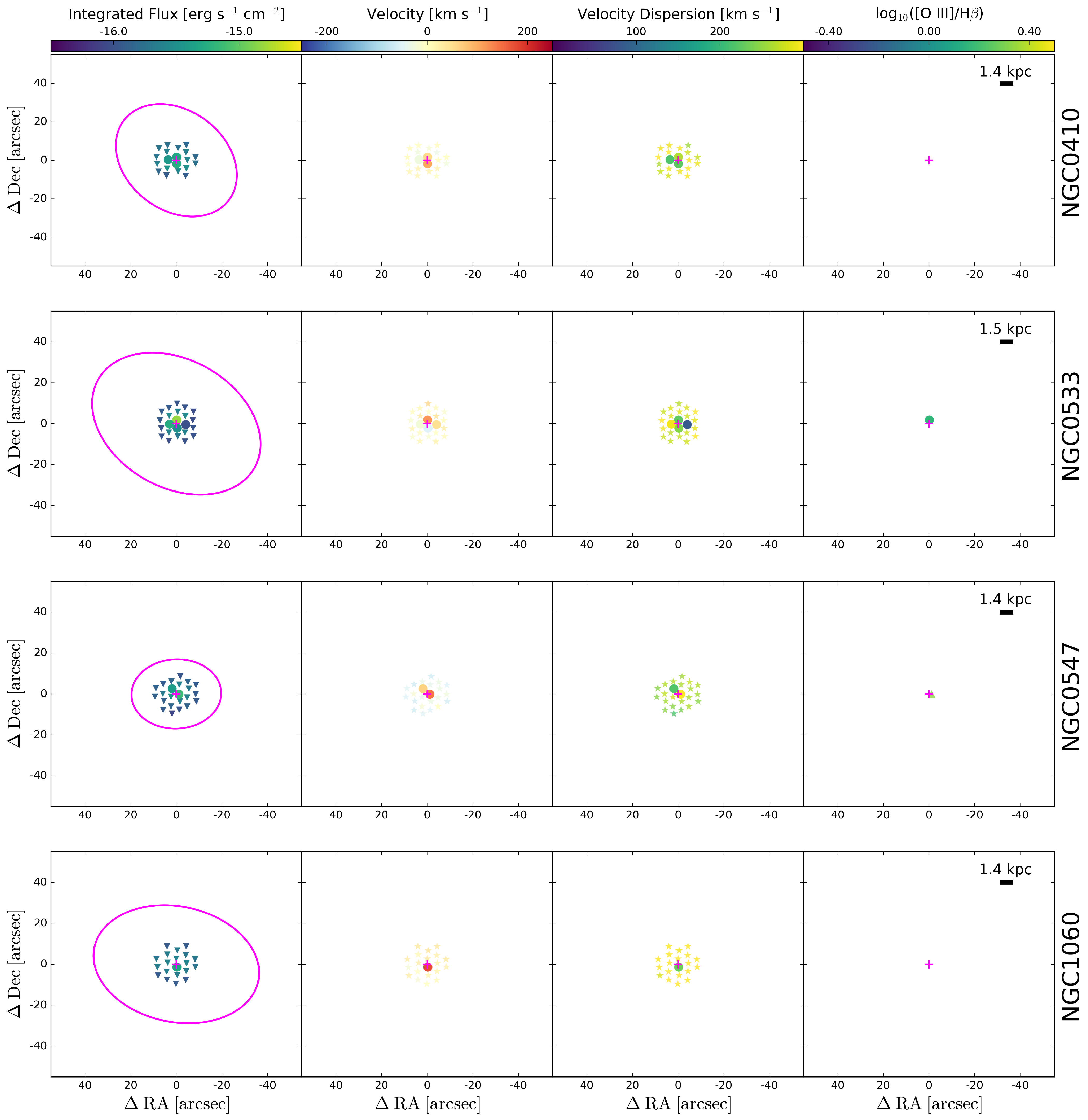}
\end{center}
\caption{Two-dimensional maps of various quantities for MASSIVE galaxies with detected warm ionized gas. From left to right: [\ion{O}{2}] integrated flux, gas and stellar velocity, gas and stellar velocity dispersion, and [\ion{O}{3}]/H$\beta$ integrated flux ratio. The plotting conventions, symbols and colors are the same as described in \autoref{fig:portraits}.}
\end{figure*}

\begin{figure*} 
\begin{center}
\includegraphics[width=\hsize]{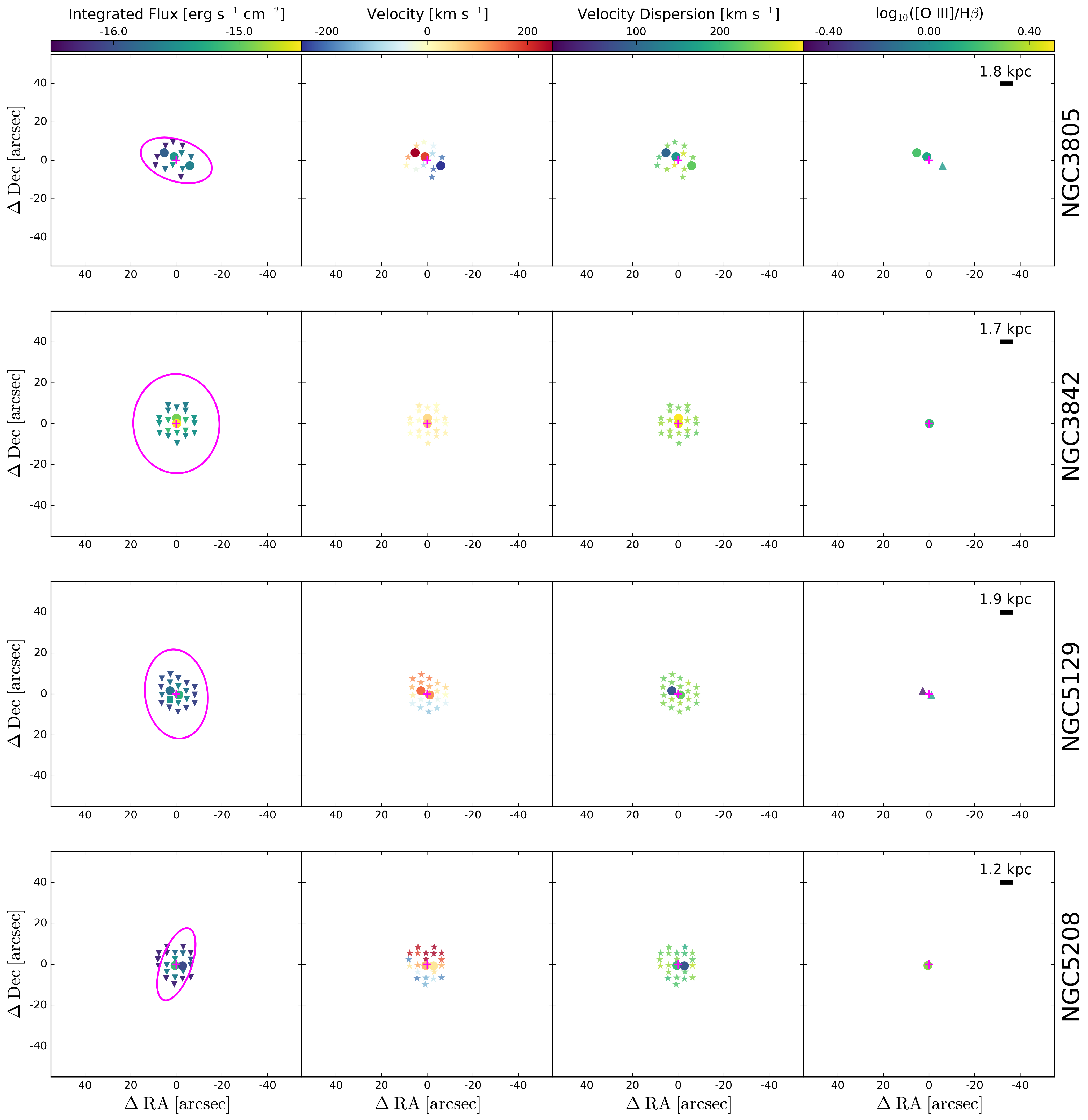}
\end{center}
\caption{Two-dimensional maps of various quantities for MASSIVE galaxies with detected warm ionized gas. From left to right: [\ion{O}{2}] integrated flux, gas and stellar velocity, gas and stellar velocity dispersion, and [\ion{O}{3}]/H$\beta$ integrated flux ratio. The plotting conventions, symbols and colors are the same as described in \autoref{fig:portraits}.}
\end{figure*}

\begin{figure*} 
\begin{center}
\includegraphics[width=\hsize]{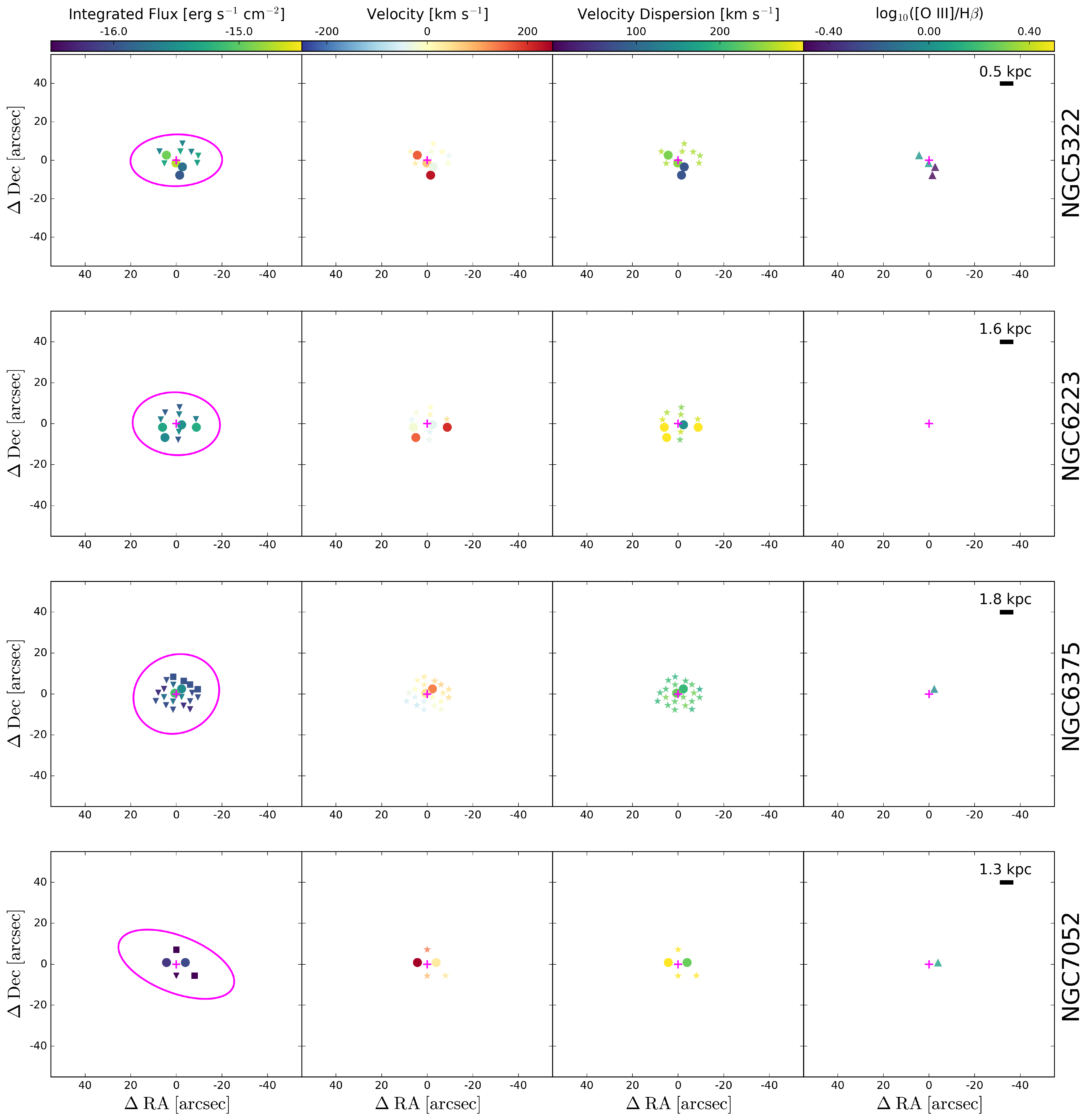}
\end{center}
\caption{Two-dimensional maps of various quantities for MASSIVE galaxies with detected warm ionized gas. From left to right: [\ion{O}{2}] integrated flux, gas and stellar velocity, gas and stellar velocity dispersion, and [\ion{O}{3}]/H$\beta$ integrated flux ratio. The plotting conventions, symbols and colors are the same as described in \autoref{fig:portraits}.}
\end{figure*}

\begin{figure*} 
\begin{center}
\includegraphics[width=\hsize]{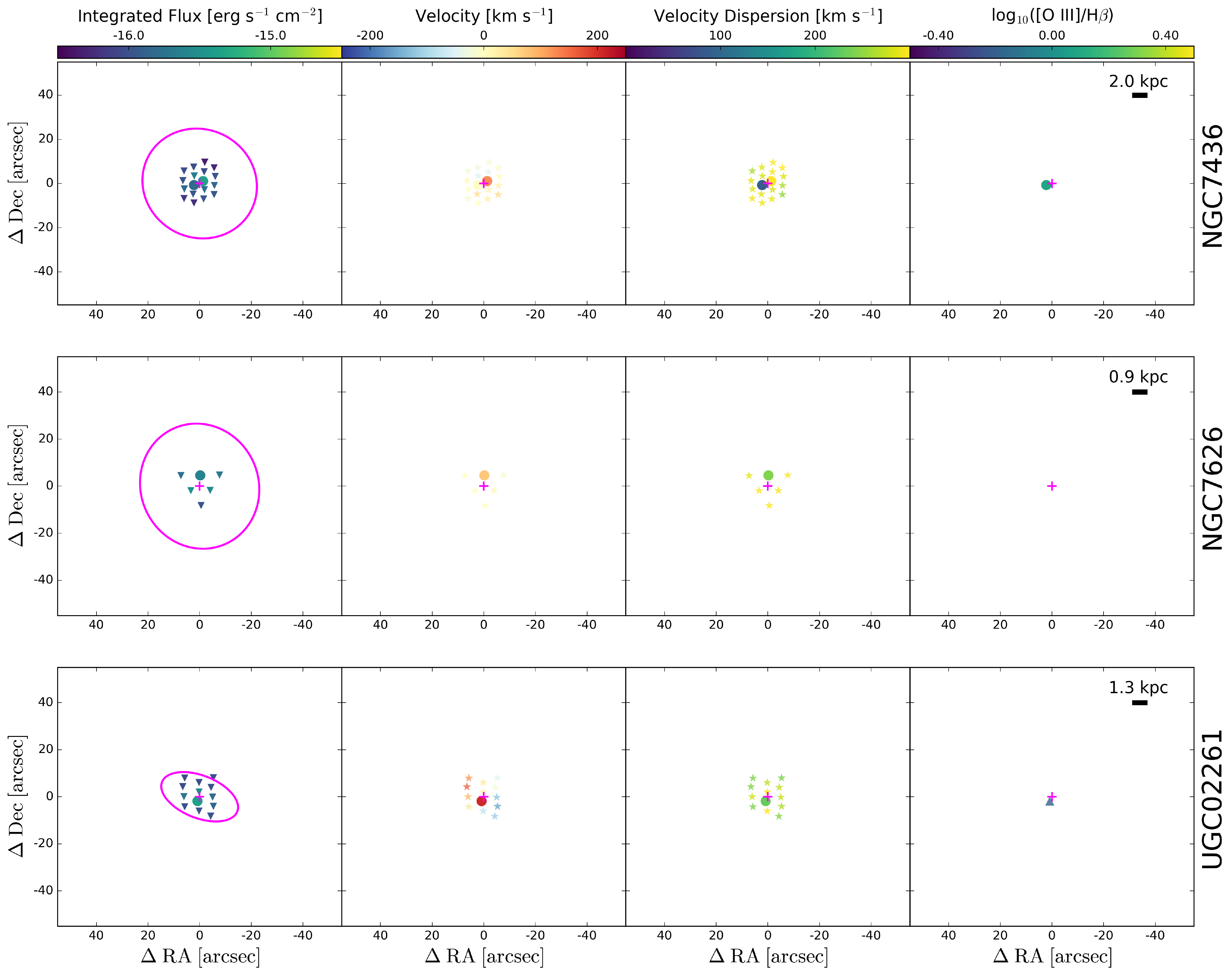}
\end{center}
\caption{Two-dimensional maps of various quantities for MASSIVE galaxies with detected warm ionized gas. From left to right: [\ion{O}{2}] integrated flux, gas and stellar velocity, gas and stellar velocity dispersion, and [\ion{O}{3}]/H$\beta$ integrated flux ratio. The plotting conventions, symbols and colors are the same as described in \autoref{fig:portraits}.}
\end{figure*}

\end{document}